\documentstyle[aps,amssymb,epic,preprint,curves]{revtex}

\def\mc{{\cal M}}

\def\tr{{\rm Tr\, }}
\def\c{{\Bbb C}}

\def\dd{\hbox{\kern0.3em/\kern-0.7em /\kern0.5em}}
\def\lie#1{{\rm Lie}\left( #1 \right)}

\def\df{$D-$flat }

\def\a{\alpha}
\def\b{\beta}

\def\n{\nu}

\def\t{\theta}
\def\hp{\hat{\phi}}
\def\hw{\hat{W}}

\def\q{\tilde{Q}}

\def\s#1{{\Bbb S}_{#1}}
\def\S#1{\Sigma_{#1}}
\def\t#1{{\Bbb T}_{#1}}
\def\1#1{{\Bbb N}_{#1}}
\def\n#1{{\Bbb N}_{#1}}
\def\ws{{\Bbb S}_W}

\def\ssqr#1#2{{\vbox{\hrule height #2pt
\hbox{\vrule width #2pt height#1pt \kern#1pt\vrule width #2pt}
\hrule height #2pt}\kern- #2pt}}

\newcommand{\drawsquare}[2]{\hbox{%
\rule{#2pt}{#1pt}\hskip-#2pt
\rule{#1pt}{#2pt}\hskip-#1pt
\rule[#1pt]{#1pt}{#2pt}}\rule[#1pt]{#2pt}{#2pt}\hskip-#2pt
\rule{#2pt}{#1pt}}

\newcommand{\Yfund}{\raisebox{-.5pt}{\drawsquare{6.5}{0.4}}}
\newcommand{\Yasymm}{\raisebox{-3.5pt}{\drawsquare{6.5}{0.4}}\hskip-6.9pt%
        \raisebox{3pt}{\drawsquare{6.5}{0.4}}}

\begin{document}
\tightenlines


\preprint{\vbox{
\hbox{hep-th/0006188}
}}
\title{Higgs Mechanism and Luna Strata in ${\cal N}=1$ Gauge Theories}
\author{Gustavo Dotti}
\address{FaMAF, Universidad Nacional de C\'ordoba,\\
Ciudad Universitaria, 5000, C\'ordoba, ARGENTINA}
\date{June 2000}

\maketitle

\begin{abstract}
The classical moduli space $\mc$ of a supersymmetric gauge theory with trivial
superpotential can be stratified according to the unbroken gauge
subgroup at different vacua.
We apply known results about this stratification  to obtain
 the $W \neq 0$ theory classical moduli space $\mc ^W \subset \mc$,
working entirely with the composite gauge invariant operators $\hp$
that span $\mc$,
assuming we do not know their elementary matter chiral field content.
In this construction, the patterns of gauge symmetry breaking
of the $W \neq 0$ zero theory are   determined,
Higgs flows in these theories show important differences
from the $W=0$ case.
The methods here introduced  provide an alternative way
 to construct tree level superpotentials
that lift all classical flat directions
leaving a candidate theory for
dynamical supersymmetry breaking, and
 are  also useful to identify heavy composite fields  to integrate
out from effective superpotentials when the elementary field content
of the composites is unknown.
We also show how to
recognize the massless singlets after Higgs mechanism  at a vacuum
 $\hp \in \mc^W$
among the moduli $\delta
\hp$ using the stratification of $\mc$, and establish conditions under
which the space of non singlet massless fields after Higgs mechanism
(unseen as moduli $\delta \hp$) is null.
A small set  of theories
 with so called ``unstable'' representations of
the complexified gauge group is shown to  exhibit
unexpected properties regarding the dimension of their
moduli space, and the presence of non singlet massless fields after
Higgs mechanism  at all of their vacua.
\end{abstract}
\pacs{11.30.Pb; 11.15-q; 11.15.Kc; 12.60.Jv}


\section{Introduction} \label{intro}

The construction of the classical
moduli space $\mc$ of a supersymmetric gauge theory with trivial superpotential
is well known~\cite{plb,gatto,gatto2,luty}: starting from the elementary chiral
matter fields $\phi \in \c^n$, a basic set $\hp^i(\phi), i=1,...,s$ of
holomorphic gauge invariant composites is
obtained. Generically, the basic invariants are constrained, there
are  polynomials
$p_{\a}(\hp)$ such that $p_{\a}(\hp(\phi))$ vanishes identically.
The classical moduli space $\mc$, defined
to be the set of $\df$ points mod the gauge group action, can be
shown to be parameterized by
the subset of $\c^s$ defined by the constraints among
the invariants,  $\mc = \{ \hp \in \c^s | p_{\a}(\hp)=0 \}$~\cite{plb,luty}.
It is  worth recalling at this point
 that $\mc$ agrees with the  {\em quantum}
moduli space of the theory if the Dynkin index of the gauge group representation
on the elementary field space  is greater than the index of the adjoint
representation~\cite{sc}.
$\mc$ also has a geometrical interpretation~\cite{plb,gatto}:
if $G^c$ is the complexification of the gauge group $G$, then
$G^c$ is non-compact and
some of the  $G^c$ orbits in $\phi$ space $\c^n$ are not closed.
$\mc$  is shown to parameterize the set of {\em closed} $G^c$ orbits,
denoted $\c^n//G$ to distinguish it from orbit  space $\c^n/G$.
 The  relation  $\mc = \c^n//G$
is due to the fact that there is
precisely one $G$ orbit of \df points per closed $G^c$ orbit, and no
\df points in non-closed $G^c$ orbits~\cite{plb,gatto}.\\
Now suppose we add a tree level superpotential $W(\phi)$.
To ensure gauge invariance, we must have $W(\phi)=\hw(\hp(\phi))$,
where $\hw: \c^s \to \c$ is an arbitrary function on the basic invariants
(the distinction of the superpotential $\hw$ as a function of the basic invariants
from the superpotential $W$ as a function of the elementary fields
is crucial in what follows.)
 The classical moduli space $\mc ^W \subset \mc$
of the theory with the added superpotential is
the image under  $\pi: \phi \to \hp(\phi)$
of the set $dW=0$ of $F-$flat points in $\c^n$.
In~\cite{luty} it is shown that $\mc^W \subset \mc \subseteq \c^s$
can be obtained by
adding to the algebraic constraints $p_{\a}(\hp)=0$ among
the invariants the gauge invariant constraints resulting
from $dW=0$.
A natural question to ask
is  the following:
suppose we are given $\mc$ (i.e., the number $s$ of basic invariants
and the constraints $p_{\a}: \c^s \to \c$) and  $\hw(\hp)$,
 but we {\em do not
know} the elementary field composition $\hp(\phi)$ of the basic
invariants (in particular,
we do not know $W(\phi)=\hw(\hp(\phi))$).
Is it possible to construct $\mc^W$ from this information?
 This would give us
what we may  call a ``low energy description'' of $\mc^W$,
since only the composite fields are involved in the construction.
At first sight, we may  think  that
knowledge of the constraints linking the basic invariants $\hp$,
the ones
that define $\mc$, is enough.
For example, if $\hw = m \hp^1$ is a mass term and we know the constraints
linking $\hp ^1$ to the other composite superfields $\hp $, we may think
we should be able to deduce which composite superfields
are made heavy by $\hw$.
Unfortunately, this is not the case, a ``low energy"
 description is not possible unless further input is given.
The following is probably the simplest example
illustrating  this fact:
consider an $SO(N)$ theory with two flavors of vector fields,
$\{ \phi \} = \{Q_i^{\a}, \a=1,...,N, i=1,2\}\simeq  \c^{2N}$.
The basic invariants are $\{ \hp \} = \{ S_{ij} \equiv Q^{\a}_i
Q^{\a}_j \}$, and $\mc = \{ S_{ij} \} \simeq \c^3$, as there are
no constraints among the invariants.
Although the directions $S_{11}, S_{12}$ and $S_{22}$ in $\c^3$ are
completely equivalent, $\mc^W=\{ (0,0,0) \}$ if $\hw = mS_{12}$, whereas
$\mc^W = \{(S_{11},0,0)\} \simeq \c^1$ if $\hw =mS_{22}$. The example
shows that knowledge of the invariants $\hp$, their constraints, and
$\hw(\hp)$ is not enough to obtain $\mc^W$,
an extra piece of information is required.
The zero superpotential moduli space  $\mc$
can be stratified according to the conjugate class  $(H)$ of
the unbroken gauge subgroup $H \subseteq G$ at each vacuum.
The stratum $\S {(H)} \subset \mc$
contains all vacua with unbroken gauge subgroup
conjugate to $H$.
It turns out that the
 stratification $\mc = \cup_{(H)} \Sigma_{(H)}$ is
 precisely the extra piece of
information required to accomplish the desired low energy description.
The relation between the stratification of
$\mc$ and the low energy construction of $\mc^W$ comes from the
equality  $\mc^W \cap \Sigma _{(H)} =
\{ \hp \in \Sigma _{(H)} | d \hw_{(H)} (\hp) = 0 \}$,
$\hw_{(H)}$ being the restriction ${\hw}_{|_{\S {(H)}}}$
of $\hw$  to $\Sigma_{(H)}$.
 $\mc^W \subset \mc$ can be constructed in steps by
finding the stationary points of the restriction
of $\hw$ to $\Sigma _{(H)}$,
one
stratum at a time.
This useful fact, pointed out  in \cite{plb}, follows from 
results of Luna~\cite{luna}, Abud and Sartori~\cite{as},
Procesi and  Schwarz~\cite{plb,schwarz1}.
 In this paper we elaborate further
on these results  and obtain an algorithm to construct
$\mc^W$ which, in some cases,
 saves us the job of looking
for critical points in every stratum, but only on some carefully
chosen ones.
These techniques are  applied to recognize heavy composites
(of unknown elementary field content) to integrate out from
an effective superpotential $W_{eff}(\hp)$~\cite{out,susy}.
They are also used to
construct tree level superpotentials $\hw$ that lift all non-trivial
flat directions, reducing the classical moduli space to a point.
In all cases the input is the stratification of $\mc$, where the
calculations are performed, the
composition $\hp(\phi)$ of the basic invariants in terms of the elementary fields
is not required.
Theories lifting all non trivial flat directions
are interesting as candidates for
dynamical supersymmetry breaking~\cite{dsb}.
We finally use the results in~\cite{plb}
to investigate   the relationship (in the classical theory)
between the massless modes $\delta \phi$ at a vacuum $\phi$ 
in unitary gauge, and the moduli $\delta \hp$
obtained by linearizing at $\hp(\phi)$
the constraints among the $\hp$'s.
The expected isomorphism between these two sets
holds (in most theories) only at the
so called principal stratum $\Sigma_{(G_P)}$,  where the gauge
group $G$ is maximally broken.
Yet, some exceptional theories are found for which the isomorphism does not
hold even a the principal stratum. This is the same set of theories for which
the equation $\text{dim } \mc = \text{ dim } \{\phi\} -
(\text{ dim}_{\Bbb R}\, G
- \text{ dim}_{\Bbb R}\, G_P)$ does not hold\footnote{dim denotes
complex dimension, whereas $\text{dim}_{\Bbb R}$ means real
dimension, then dim~$G^c = \text{dim}_{\Bbb R}\, G$.}, they are
 characterized by the fact
that the bulk of the configuration space $\{ \phi \} \simeq \c^n$ is filled with
non closed orbits of the complexification $G^c$, case in which
the $G^c$ action on $\phi$ space is
termed ``unstable".
Since the $G$ representation on $\c^n$ must be anomaly free, most
anomaly free representations are real, and real representations are stable,
unstable theories are rare. \\
The paper is organized as follows.
In Section~\ref{ls} we introduce the stratification of $\mc$
and an order relation between strata.
The important results of Luna, Procesi  and Schwarz
are integrated in Theorem~I in Section~\ref{lt},
examples are given in Section~\ref{exs}.
In Section~\ref{app} we apply Theorem~I to a number of problems.
The low energy construction of $\mc^W$,
is treated in Section~\ref{lec}, in Section~\ref{comp}
we show the usefulness of breaking  $\mc^W$ up into
its irreducible components,
and study the patterns of gauge symmetry breaking in $W \neq 0$
theories, the problem of identifying heavy composites,
and that  of constructing superpotentials that lift all non trivial vacua.
In Section~\ref{mf} we study the relation between massless fields
after Higgs mechanism (MFHM) at a
vacuum $\hp_0 \in \mc^W$ and the space of moduli tangent to $\mc^W$ at $\hp_0$.
A number  of examples is given, many of them were constructed to
illustrate the   subtleties involved
in the  given results.
Section~\ref{conc} contains the conclusions.
We defer to Appendix~\ref{appen}  some technical aspects in the derivation of the
results in Section~\ref{app}.

\section{Luna's Stratification of the Moduli Space} \label{ls}

Let $\{\phi\}\simeq \c^n$ be the set of matter chiral fields of
a supersymmetric gauge theory with gauge group $G$ and
zero superpotential, $\hp^i(\phi), i=1,...,s$ a basic set of holomorphic
$G$ invariant operators, $p_{\a}(\hp)=0, \a=1,...,l$ the algebraic constraints
among the basic invariants. The moduli space of the theory is
$\mc = \{ \hp \in \c^s | p_{\a}(\hp)=0 \}$. This means that
for every  $\hp_0$ satisfying $p_{\a}(\hp_0)=0$
there is precisely one $G$ orbit $G \phi_0$ of \df points
satisfying $\hp(\phi_0)=\hp_0$. Note that $G \phi$ denotes the
$G$ orbit through $\phi$, whereas $G_{\phi}$ denotes  the unbroken
gauge subgroup at $\phi$. Since points in the same $G$ orbit have
conjugate little groups, $G_{g\phi} = gG_{\phi}g^{-1} \forall g \in G$,
a conjugate class $G_{\hp_0}$ can be associated to $\hp_0 \in \mc$, namely,
$(G_{\hp_0}) \equiv (G_{\phi_0})$, where $\phi_0$ is any \df point
satisfying $\hp(\phi_0)=\hp_0$. The definition makes sense since
any two \df points $\phi_0, \phi_1$ satisfying $\hp(\phi_0)=\hp_0=\hp(\phi_1)$
 are $G$ related. A  stratum $\S {(H)}$ is
the set of $\hp$'s in $\mc$ satisfying $(G_{\hp})=(H)$, $\mc =
\cup_{(H)}\S {(H)}$ is the disjoint union of its strata. The strata are
complex manifolds of different dimensions, $\mc$ instead is an
{\em algebraic set}~\cite{clo}, the zero set of a family of polynomials.
The tangent space at a point $x \in X$, $X$ an algebraic set or a complex
manifold, is denoted $T_xX$. For an algebraic set
$X = \{ x \in \c^s | p_{\a}(x)=0, \a=1,...,l\}$,
$T_xX$ is defined to be the kernel of the matrix
$\partial p_{\a}/ \partial x^i(x)$,
i.e., the $\delta x'$s allowed by the linearized constraints.\footnote{Note
however that  different sets of polynomials define the same algebraic
set, $\{p_{\a}\}$ must be chosen such that any polynomial $p$
vanishing on $X$ admits an expansion $p(x) = \sum_{\a} q^{\a}(x) p_{\a}(x)$
with polynomials $q^{\a}$~\cite{clo}. Otherwise,
the span of the linearized constraints may be larger than
the tangent space. As an example,
the line $x_2=0$ in $\c^2 =
\{ (x_1,x_2) \}$ can also be defined as the zero set of the
polynomial $(x_2)^2=0$, but this second choice leads to a wrong
definition of tangent space.} Generically, the dimension of the tangent
space of an algebraic set may change from point to point. If $X$ is
an algebraic set satisfying dim~$T_xX=n \; \forall x \in X$, then
$X$ is a complex manifold of dimension $n$~\cite{kodaira}.
The projection map $\pi: \phi \to \hp(\phi)$ sends $\c^n$ onto $\mc$.
Its differential at $\phi$, $\pi'_{\phi}: T_{\phi}\c^n \simeq \c^n \to
T_{\pi(\phi)}\mc$ relates the  $\delta \phi$ at $\phi$ with
the moduli $\delta \hp$ at $\hp$, $\pi'_{\phi}:
\delta \hp \to \partial \hp^i(\phi)
/ \partial \phi^j \delta \phi^j$.
An order relation
can be introduced in the set of isotropy classes, we say that $(H_1)
< (H_2)$ if $H_1$ is conjugate to a subgroup of $H_2$. This order
relation is {\em partial}, it is not true that given any two classes
$(H_1) \neq (H_2)$ either $(H_1) < (H_2)$ or $(H_1) > (H_2)$, there are
unrelated classes. The partial order relation among conjugate classes induces
a partial order relation among the  strata: $\S {(H_1)} >
\S {(H_2)}$ whenever $(H_1) < (H_2)$.

\subsection{A theorem on the stratification of the moduli space}
\label{lt}

The important results in \cite{plb,im} are the
following (see also~\cite{luna,as,schwarz1}):\\

\noindent
{\bf Theorem I:}
\begin{itemize}
\item[{\bf (a)}] There are only finitely many strata of $\mc$.
The strata
are complex manifolds, their closures are algebraic subsets of
$\mc$.
\item[{\bf (b)}] The closure of $\S {(H)}$ is
\begin{equation} \label{closure} 
\overline{\S {(H)}} = \bigcup_{(L) \geq (H)} \S {(L)},
\end{equation}
i.e., the boundary of $\S {(H)}$ is the union of the strata that are
strictly smaller than $\S {(H)}$.
\item[{\bf (c)}] There is a unique minimal isotropy class $(G_P)$, called
{\em principal isotropy class}, $\S {(G_P)}$ is called principal
stratum. 
$(G)$ is a unique maximal isotropy class.
\item[{\bf (d)}] Assume $\phi$ is  \df and let
$\t {\phi} \equiv \lie {G^c}
{\phi} \simeq T_{\phi}G^c \phi$,
the tangent at $\phi$ of the $G^c$ orbit through $\phi$.
$\t {\phi} \subset \c^n$ is a $G_{\phi}$ invariant subspace ,
and it has a $G_{\phi}$ invariant complement ${\t {\phi}}^{\perp}$.
The theory with gauge group $G_{\phi}$ and
matter content ${\t {\phi}}^{\perp}$  is called {\em slice representation}.
The stratification of the moduli space
of the
slice representation
contains precisely the $(H) \leq (G_{\phi})$ classes of the original theory.
\item[{\bf (e)}]
Let $\s {\phi} \subseteq {\t {\phi}}^{\perp}$ be the subspace of
$G_{\phi}$ singlets,
$\n {\phi}$ a $G_{\phi}$ invariant complement of $\s {\phi}$ in
${\t {\phi}}^{\perp}$, then $\c^n = \t{\phi} \oplus
\s {\phi} \oplus \n {\phi}$.
The differential $\pi'_{\phi}$ of the projection map $\pi$ at $\phi$
has kernel
 $\t {\phi} \oplus \n {\phi}$, its rank is $T_{\pi(\phi)} \S {(G_{\phi})}$,
the tangent to the stratum through $\pi(\phi)$.
\item[{\bf (f)}]
Assume  the \df point $\phi$ satisfies  $\pi(\phi) \in  \S {(G_P)}$.
Then $\n {\phi} = \{0\}$  if and only if the $G^c$ representation on $\c^n$
 is stable.
 If the representation is unstable, the theory with gauge group $G_P$
and matter content $\n {\phi}$ (i.e., the slice theory without the singlets)
has no holomorphic $G_P$ invariants.
\end{itemize}
Some explanations are in order. Regarding point (c)
 note that in a partially ordered
set $U$ there may be more than one maximal element. Generically, there is
a subset $M \subset U$ of maximal elements. Any two elements in $M$ are
unrelated under $<$, whereas $m>p$ for all $m \in M, p \in U \setminus M$.
 Analogously, there is a subset  of
minimal elements of $U$ . Regarding  point (d)
note that the ``slice representation" is
just the supersymmetric gauge theory obtained by
Higgs mechanism at energies below the masses of the broken generators.
An interesting observation in~\cite{im} is that $G_{\phi}$ determines
entirely the slice representation, i.e., there  cannot be two
different \df points leading to theories with the same
(class of) $G$ subgroup as gauge group but
  having  different matter content. This is a consequence of
the following identity of direct sums of $G_{\phi}$ representations
($\rho$ stands for the $G$ representation on $\{ \phi \} = \c^n$, whereas
$\rho_{|_{H}}$ means its restriction to the $G$ subgroup $H$.)
\begin{equation}
\s {\phi} \oplus \n {\phi} \oplus (Ad \; G)_{|_{G_{\phi}}} = \
\rho_{|_{G_{\phi}}} \oplus Ad \; G_{\phi},
\end{equation}
Theorem~I.c-d guarantees that {\em any} pattern of symmetry breaking from
$G$ to subsequently smaller $G$ subgroups lead to the
theory with maximally broken gauge subgroup $G_P$. According
to Theorem~I.f this theory contains only $G_P$ singlets, except
in those cases where $\rho$ is unstable. As explained above,
the complexification $G^c$ of the gauge group is non-compact, and some
of its orbits are not closed.  $\rho$ is said to be unstable if there is a
$G^c$ invariant
subset of $\c^n$, open in the Zariski topology,
 containing only non-closed $G^c$ orbits.
The Zariski topology on $\c^n$~\cite{clo} is the one whose
closed sets are algebraic sets, i.e., zeroes of a family of polynomials,
it is coarser than the usual $\c^n \simeq {\Bbb R}^{2n}$ topology.
This topology is useful in studying representations of algebraic
groups, of which the complexification $G^c$ of the compact Lie
group $G$ is an example. Zariski open subsets of a vector
space $\c^n$ are (Zariski) dense, we may therefore
view unstable theories as those for which the bulk of the elementary field
space $\c^n$ is filled with non-closed $G^c$ orbits, i.e.,
orbits without \df points. It was shown in~\cite{schwarz1}
that if the $G$ representation $\rho$ on $\c^n$ is real
then it is stable. As physical theories  must be free of
gauge anomalies, and most anomaly free representations are real,
unstable supersymmetric gauge theories are rare.
In fact, the only unstable theories based on
a simple gauge group are
$SU(2N+1)$ with $\Yasymm + (2N-3) \overline{\Yfund}, N \geq 2,$ and
$SO(10)$ with a spinor. These theories exhibit some curious properties,
as we will see.\\
Note from (b-c) that $\mc = \overline{\S {(G_P)}}$, this leads to the definition
dim~$\mc=$~dim~$\S {(G_P)}$ (in agreement with the standard definition
of dimension of an irreducible algebraic set~\cite{clo}).
The dimension of an algebraic set may change from point to
point, generically there are {\em singular points}
$\hp \in \mc$ at which dim~$T_{\hp} \mc > \text{dim }\mc$,
 they belong to smaller
strata.  As stressed in~\cite{luty}, however,
it is not true that all vacua $\hp$ satisfying $(G_{\hp})> (G_P)$ are
singular, a trivial counterexample being offered by those theories with
unconstrained basic invariants, for which all points
of $\mc \simeq \c^s$ are non-singular, including those with enhanced
gauge symmetry. \\
From Theorem~I we  can  show that
\begin{equation} \label{int}
\S {(H')} \cap \overline{\S {(H)}} \neq \emptyset \Rightarrow
\S {(H')} \leq \S {(H)} \;\; \text{ (equivalently }
\S {(H')} \subseteq \overline{\S {(H)}}).
\end{equation}
This is proved by taking $\phi \in \S {(H')} \cap \overline{\S {(H)}}$,
then $(G_{\phi}) = (H')$ and also, using Theorem~I.b,
$(G_{\phi}) \geq (H)$, from where equation~(\ref{int})  follows.
Another straightforward consequence
of the theorem is that, for stable actions (only!), $\text{dim } \mc = n -
\text{ dim  } G^c + \text{dim }  {G_P}^c $.
This is proved by   picking a \df point $\phi$
satisfying $\pi(\phi) \in \S {G_P}$. We have the following (in)equalities
from (b,e) of Theorem~I:\footnote{For a
\df point $\phi$,  ${G^c}_{\phi} = {G_{\phi}}^c$~\cite{gatto}, then the  complex
dimension  $\text{dim } {G^c}_{\phi}$ equals the real dimension
 $\text{dim}_{\Bbb R}\, G_{\phi}$.
In particular, if  $\pi(\phi)$ is
 in the principal stratum,
 $\text{dim } {G^c}_{\phi} =  \text{dim}_{\Bbb R}\, G_{\phi}
= \text{dim}_{\Bbb R} \, G_P = \text{dim } {G_P}^c.$}
$\text{dim } \mc \equiv \text{dim }
\S {(G_P)}  = \text{rank } \pi'_{\phi}
 = n - \text{ dim ker } \pi'_{\phi} = n - \text{ dim } \t {\phi} -
\text {dim } \n {\phi} =
n -(\text { dim } G^c - \text{ dim }  {G_P}^c) -
\text {dim } \n {\phi} \leq 
n -(\text { dim } G^c - \text{ dim }   {G_P}^c)$.
According to Theorem~I.f, equality holds only if $\rho$ is stable.
For unstable theories the dimension of $\mc$ is smaller than the
expected value $ n -
\text{ dim  } G + \text{dim}_{\Bbb R}  G_P$, this is
consistent with the statement above that ``the bulk of $\phi$ space"
(a Zariski dense subset) contains no \df point. Unstable theories
{\em do} have $G^c$
orbits of dimension equal to $n - \text{dim }\mc >
\text{dim } G^c - \text{dim } {G_P}^c$~\cite{ela},
 however, there is no \df point in
these highest dimensional orbits. In other words, unstable theories are
characterized by the impossibility of breaking $G^c$ to the smallest isotropy
$G^c$ subgroup by a \df point.

\subsection{Examples} \label{exs}

In the following, we will arrange partially ordered sets $U$ in columns
in this way: the first column (from left to right) contains
the subset $C_ 1 \subset U$ of maximal elements in $U$,
the second column contains  the subset $C_2$ of  maximal
elements in $U \setminus C_1$, the third column $C_3$ contains
the maximal elements in $U \setminus (C_1 \cup C_2)$, and so on.
We will also draw a line linking the elements in adjacent
columns which are related under $<$. Note that, by construction,
any element in $C_{i+1}$ is smaller than at least one element in
$C_i$. Note also from Theorem~I.c that if $U$ is the set of strata $\S {(H)}$
or conjugate classes $(H)$, then the first and last column contain a single
element. For totally ordered sets there is a single entry per column.\\

Our first example is a  theory with a smooth moduli space
$\mc \simeq \c^s$ and  totally ordered strata.\\
\noindent
{\it Example \ref{exs}.1:} Consider $F$ flavor, $N$ color SQCD 
with quarks $Q^{\a}_i$ and antiquarks $\q_{\b}^j$, $\a,\b = 1,...,N$;  
$i,j = 1...F$, $F<N$.
The basic invariants are $M^j_i = \q ^j_{\a} Q^{\a}_i$, they are 
unconstrained and so $\c^{F^2} \simeq \mc = {\Bbb M}^F $, 
the set of $F \times F$ complex matrices.  
The classical global non-R symmetries are $K = U(F)_Q \times
U(F)_{\q}$. A generic \df point can be 
$G \times K$ rotated onto
\begin{equation} \label{dfqcd}
Q^{\a}_i = {({\q}^{\dagger})}_{\beta}^j = \left( \begin{array}{cc} V & 0 \\ 0 & 0 
\end{array} \right), \hspace{1cm} 
V = \left( \begin{array}{ccccc} 
v_1 & 0 & 0 & \cdots & 0\\
0 &  v_2 & 0 & \cdots &0\\
\vdots & \vdots &  & & \vdots\\
0 & 0 & \cdots & v_{r-1} & 0 \\
0 & 0 & \cdots & 0 & v_r \\ 
\end{array} \right), \;\;\ v_i \neq 0, \;\; r \leq F. 
\end{equation}
As isotropy $G$ subgroups are $K$ invariant and $G$ conjugate
we only need 
 consider the \df points~eq.(\ref{dfqcd}) to obtain Luna's 
stratification of $\mc$. The unbroken $G$ subgroup  at $(Q,\q)$ 
of equation~(\ref{dfqcd})  is  $SU(N-r)$
($SU(1)$ meaning the trivial group). There are 
$F+1$ strata, $\S {SU(N-r)}$, $r=0,1,...,F$, and there is a {\em complete} 
order relation 
$\S {SU(N)} < \S {SU(N-1)}~<~\cdots~< \S {SU(N-F)}$,
then we arrange the strata as
$$\S {SU(N-F)} - \S {SU(N-F-1)} - \cdots - \S {SU(N-1)} - \S {SU(N)}.$$
From~(\ref{dfqcd}) follows  that $\S {SU(N-r)}$ is the 
set of $K$ orbits of points
$M = \text{diag} (|v_1|^2,...,|v_{r}|^2,0,...,0),\; |v_i| \neq 0$, 
which is  the set ${\Bbb M}^{F}_r$ of rank $r$ complex $F \times F$ matrices.
The determinantal variety~\cite{harris} ${\Bbb M}^F_{\leq r}$ of $F \times F$ 
matrices of rank less than or equal to $r$ is the algebraic set 
\begin{equation} \label{detva}
{\Bbb M}^F_{\leq r}  = \{ M \in {\Bbb M}^F | 
M^{[j_1}_{i_1}M^{j_2}_{i_2} \cdots M^{j_{r+1} ]}_{i_{r+1}} = 0 \}.
\end{equation}
As ${\Bbb M}^F_r = {\Bbb M}^F_{\leq r} \setminus {\Bbb M}^F_{\leq r-1}$,
equation (\ref{detva}) defines the smallest Zariski closed (i.e., algebraic)
set containing
${\Bbb M}^F_{r}$, i.e., ${\Bbb M}^F_{\leq r}=\overline{{\Bbb M}^F_r}$.
This verifies Theorem~I.b:
$\overline{\S {SU(N-r)}} = \cup_{j \leq r} \S {SU(N-j)}$.
It is instructive to see what
the tangent space $T_M  {\Bbb M}^F_{\leq r}$ is
(for an alternative derivation see~\cite{harris}).
As the equations defining ${\Bbb M}^F_{\leq r}$ in~(\ref{detva})
satisfy the requirement in footnote~2,
the tangent space
at $M$ of ${\Bbb M}^F_{\leq r}$ is obtained by linearizing~(\ref{detva}),
\begin{equation} \label{tanqcd}
 T_M {\Bbb M}^F_{\leq r} = \{ \delta M \in {\Bbb M}^{F} :
M^{[j_1}_{i_1}M^{j_2}_{i_2} \cdots M^{j_r}_{i_r} 
\delta M^{j_{r+1} ]}_{i_{r+1}} = 0 \}.
\end{equation}
To understand the condition eq.~(\ref{tanqcd}) contract 
 $M^{[j_1}_{i_1}M^{j_2}_{i_2} \cdots M^{j_r}_{i_r} 
\delta M^{j_{r+1} ]}_{i_{r+1}} = 0 $ with $r+1$ linearly independent 
vectors
$t^{i_k}_{(k)}, k=1,...,r+1$. If rank~$M < r$
 at least two of the $t$ vectors belong to 
ker $M$, (\ref{tanqcd}) is trivially satisfied for any matrix $\delta M$,
$T_M {\Bbb M}^F_{\leq r} \simeq  {\Bbb M}^F$,
dim~$T_M {\Bbb M}^F_{\leq r}=F^2$.
If rank $M = r$ we  get a nontrivial
condition if we choose  the $t_{(j)}$ such that only one of
them, say $t_{(r+1)}$, belongs to the kernel of $M$. The condition
is $M^{[j_1}_{i_1}M^{j_2}_{i_2} \cdots M^{j_r}_{i_r} 
\delta M^{j_{r+1} ]}_{i_{r+1}} t_{(1)}^{i_1} \cdots t^{i_{r+1}}_{(r+1)}
= 0$, meaning that   $\delta M$ must send the kernel of $M$ onto
the rank of $M$, the dimension of the tangent space at $M$, the space
of allowed $\delta M$'s,  being
$F^2-(F-r)^2$. We conclude that $\S {SU(N-r)} = {\Bbb M}_{r}^F$ is
the subset of non singular
 points of ${\Bbb M}_{\leq r}^F = \overline{\S {SU(N-r)}}$,
the dimension
of the complex manifold $\S {SU(N-r)} = {\Bbb M}_{r}^F$ being
 $F^2 - (F-r)^2$.\\
The complexification of $G$ is 
$SU(N)^c=SL(N,\c)$, and $T \in \lie {SL(N,\c)}$ can be written as
\begin{equation}
\lie {SL(N,\c)} \ni T = \left(\begin{array}{c|c} t_1 & t_2 \\ \hline t_3 & t_4
\end{array} \right), \;\;\;\; t_4 \in \lie {GL(N-r,\c)}, \tr {t_1}
+ \tr {t_4} = 0. \end{equation}
The (Lie algebra of the) 
isotropy group ${G^c}_{(Q,\q)}={G_{(Q,\q)}}^c$
of (\ref{dfqcd}) is obtained by setting
$t_1=t_2=t_3=0$, $t_4 \in SL(n,\c)$.
We also split $Q$ and $\q$ as
\begin{equation} 
Q^{\a}_i = \left(\begin{array}{c|c} q_1 & q_2 \\ \hline q_3 & q_4
\end{array} \right), \;\;\;\; 
\q^j_{\a} = \left(\begin{array}{c|c} \tilde{q}_1 & \tilde{q}_2 \\  \hline
\tilde{q}_3 & \tilde{q}_4 
\end{array} \right), \end{equation}
where $q_1$ and $\tilde{q}_1$ are $r \times r$ blocks. 
The tangent space to the $G^c$ orbit of~(\ref{dfqcd}) is
obtained by acting with $\lie {SL(n,\c)}$ on $(Q,\q)$
\begin{equation}
\t {(Q,\q)}: \delta Q ^{\a}_i =
\left( \begin{array}{c|c} t_1 V & 0 \\ \hline
t_3 V & 0 \end{array} \right), \;\;\;\;
\delta \q ^j_{\b} = \left( \begin{array}{c|c} - V^{\dagger} t_1 &
- V^{\dagger} t_2 \\ \hline 0 & 0\end{array} \right). \end{equation}
An $SU(N-r)$  invariant complement is given by  $\n {(Q,\q)} \oplus \s {(Q,\q)}$,
where
\begin{eqnarray}
\n {(Q,\q)} &:& 
\delta Q^{\a}_i = \left(\begin{array}{c|c} 0 & 0 \\ \hline 0 & \delta q_4
\end{array} \right), \;\;\;\; 
\delta \q^j_{\a} = \left(\begin{array}{c|c} 0 & 0 \\  \hline
0 & \delta \tilde{q}_4
\end{array} \right). \\
\s {(Q,\q)} &:&
\delta Q^{\a}_i = \left(\begin{array}{c|c} 0 & \delta q_2 \\ \hline 0 & 0
\end{array} \right), \;\;\;\; 
\delta \q^j_{\a} = \left(\begin{array}{c|c} \delta \tilde{q}_1 & 0 \\ \hline
\delta \tilde{q}_3 &  0 \end{array} \right).
\end{eqnarray}
The slice representation at $(\ref{dfqcd})$ is $\n {(Q,\q)} \oplus 
\s {(Q,\q)}$, the $SU(N-r)$ theory
 with $(F-r) (\Yfund + \overline{\Yfund}) + (2Fr-r^2) {\bf 1}$, 
as is well known. 
The  configuration point $(Q,\q)$ of eq.~(\ref{dfqcd}) is sent by $\pi$  to
the following point of $\mc = {\Bbb M}^{F}$:
\begin{equation}
M = \pi(Q,\q) = \left( \begin{array}{cc} V^{\dagger} V & 0 \\0 &  0\end{array}
\right). \end{equation}
It is easily verified that $\pi'_{(Q, \q)}$ annihilates $\t {(Q,\q)}
 \oplus \n {(Q,\q)}$, 
whereas 
\begin{equation} \label{rankqcd}
\text{rank} \; \pi'_{(Q,\q)} = \pi'_{(Q,\q)} (\s {SU(N-r)}) =
\biggl\{ \delta M^i_j \in {\Bbb M}^{F} : \delta M^i_j =
\left( \begin{array}{c|c} \delta \tilde{q}_1 V & V^{\dagger} \delta q_2 \\
\hline \delta \tilde{q}_3 V & 0 \end{array} \right) \biggr\}.
\end{equation}
As $V$ is invertible, (\ref{rankqcd}) agrees with the  set of matrices
sending ker $M$ onto rank $M$, which is the tangent space  
$T_M {\Bbb M}_r^F$ at $M$ of the stratum through $M$. This verifies
Theorem~I.e.\\

\noindent
The moduli space ${\cal M}$ 
of the following example contains singular points. Its strata are 
totally ordered, and  $\S {(G_P)}$ equals the set of non-singular points
of $\mc$, a property that is not generic.

\noindent
{\it Example \ref{exs}.2:} Consider $F=N$ SQCD. \df points can be 
$G \times K$ rotated onto $Q^{\a}_i = \text{diag } \;(q_1,...,q_N)$,
$\q^{j}_{\b} = \text{diag } \;(\tilde{q}_1,...,\tilde{q}_N)$
subject to
\begin{equation} \label{cond}
|q_i|^2 - |\tilde{q}_i|^2 = c, \;\; \text{  independent of } i.
\end{equation}
The invariants are $M^j_i = Q_i^{\a} \q ^j_{\a}, B = \text{det }\; Q$, 
and $\tilde{B} = \text{ det } \q$, they satisfy
\begin{equation} \label{f=n} 
\text{det } M - B \tilde{B} = 0.
\end{equation} 
If $B = \prod_i q_i \neq 0$ or $\tilde{B} = \prod \tilde{q}_i \neq 0$, 
$G$ is completely broken. If some of the 
$q$'s are zero, then the same set of $\tilde{q}$'s must be zero, 
otherwise we get 
both $c > 0$ and $c < 0$ in equation~(\ref{cond}) .
Let $r$ be the number of zero $q$'s. 
If $r=1$, $SU(N)$ is completely broken,
rank $M = N-1$, and $B=\tilde{B}=0$.
If $r > 1$, $SU(N)$ is broken to $SU(r)$, rank $M = N-r$,
and $B=\tilde{B}=0$.
We conclude that the principal stratum is 
$\S {e} = \{ (M,B,\tilde{B}) |
B \neq 0, \text { or } \tilde{B} \neq 0,  \text{ or cofactor }  M \neq 0 \}$. 
The other strata are $\S {SU(r)} = \{ (M,B,\tilde{B}) | B=\tilde{B}=0 
\text{ and rank } M = N-r \}, r > 1$.
By linearizing eq~(\ref{f=n}) we
see that $\S {e}$ agrees with the set of non singular points of
$\mc$. The $N-1$ strata are completely ordered:
$$\S {e} - \S{SU(2)} - \cdots - \S { SU(N)}.$$

\noindent
We now present  examples where  the set of strata is
only partially ordered. \\
\noindent
{\it Example \ref{exs}.3:} Consider $G=SU(N)$ with an ($SL(N,\c)$) 
adjoint field 
$A^{\a}_{\b}$. The basic invariants are $t_j=\tr A^{j+1}, j=1,...,N-1$,
they are unconstrained and so $\mc = \c^{N-1}$. The $D-$flatness 
condition is $\tr T [A,A^{\dagger}]=0, \forall T \in SU(N)$, then  
 $[A,A^{\dagger}] \propto {\Bbb I}$, and so $[A,A^{\dagger}] 
= 0$. This implies that $A$ can be $G$ rotated onto a diagonal 
complex matrix. The residual gauge symmetry, the group of permutations 
of the diagonal entries, can be used to bring  $A^{\a}_{\b}$  to the 
following form:
\begin{equation} \label{dfadj}
A = \text{diag} \; (\stackrel{m_1}{\overbrace{v_1,v_1,...,v_1}},
\stackrel{m_2}{\overbrace{v_2,v_2,...,v_2}}, \cdots ,
\stackrel{m_j}{\overbrace{v_j,v_j,...,v_j}}), 
\end{equation}
where 
\begin{equation} \label{dfadj1}
m_1 \geq m_2 \geq \cdots m_j \geq 1,\;\; 
\sum_{k=1}^j m_k = N, \;\;
\text{and}\;\;  \sum_{k=1}^j m_k v_k = 0.
 \end{equation}
The configuration point above breaks $SU(N)$ to 
$S(U(m_1) \times U(m_2) \times \cdots \times U(m_{j-1}) \times U(m_j))$, 
(block diagonal matrices of the form diag$(g_1,...,g_j)$, $g_k \in 
U(m_k)$ and $\prod_{i=1}^j \text{det} \, g_i = 1$). In some particular 
cases this is a direct product group, for example, if $m_j=1$ 
then $S(U(m_1) \times U(m_2) \times \cdots \times U(m_{j-1}) \times U(m_j)) 
= U(m_1) \times U(m_2) \times \cdots \times U(m_{j-1})$. 
The isotropy groups are in one to one correspondence 
 with the  partitions ${\cal P}$ of $N$, a partition being  
a decomposition  $N = m_1 + m_2 + \cdots m_j$ where $m_1 \geq m_2 \geq \cdots 
\geq m_j \geq 1$. 
The partial order in the set of isotropy groups induces the 
following partial order relation in the set of partitions of $N$: 
${\cal P}_1$ is smaller than
${\cal P}_2$ if  ${\cal P}_2$  is obtained from ${\cal P}_1$  
by summing some of its terms and ordering the
resulting terms. We give some $N=5$ examples:
 $2+1+1+1=2+(1+1+1)=3+2$, then $2+1+1+1 < 3+2$, also $3+2=(3+2)=5$
then $3+2 < 5$; finally, $3+1$ and $2+2$ are unrelated. It is
easy to see that 
the partitions of $N$ (and therefore the isotropy groups and strata
of the $SU(N)$ theory with an adjoint) are totally ordered if $N=2,3$,
but only partially ordered if 
$N\geq 4$. There is exactly one point of the form~(\ref{dfadj}-\ref{dfadj1})
in a  $G$ orbit of \df points, this implies that 
$\{v_1,...,v_{j-1}\}$ can be taken as a
set of local coordinates of $\S {S(U(m_1) \times \cdots U(m_j))}$. 
In particular,  $\S {S(U(m_1) \times \cdots U(m_j))}$ has 
(complex) dimension $j-1$. Starting $N=4$ we have  distinct strata
of the same dimension. According to Theorem~I.b,
two such strata must be
unrelated under $<$, as none of them can lie in the boundary of the other one.
Write 
\begin{equation}
A^{\a}_{\b} = \left( \begin{array}{cccc} t_{11} & t_{12} & \cdots & t_{1j} \\
t_{21} & t_{22} & \cdots & t_{2j} \\ \vdots & \vdots & & \vdots \\
t_{j1} & t_{j2} & \cdots & t_{jj} \end{array} \right),
\end{equation}
$t_{ik}$ is an  $m_i \times m_k$ matrix,  $\sum_k \tr t_{kk} = 0$.
The tangent space at $(\ref{dfadj})$  breaks 
up into 
\begin{eqnarray}
\t {A} &=& \{ \delta A | \delta t_{kk} = 0, \; k=1,...,j \} \\
\label{asing} \s {A} &=& \{ \delta A | \delta t_{ij}
= \delta_{ij} a_i 
{\Bbb I}_{m_i \times m_i}, \; \sum_{i=1}^j m_i a_i = 0 \} \\
\n {A}&=& \{ \delta A | \delta t_{ij} =
\delta_{ij} t_{ii}, \tr t_{kk}=0 \; \text{for} \; 
k=1,...,j \}
\end{eqnarray}
It is readily verified that $\pi'_{A}$ annihilates $\t {A} \oplus \n {A}$.
The easiest way to see that $\pi'_{A}$ sends 
$\s {A}$ isomorphically onto $T_{\pi(A)}
\S {S(U(m_1) \times \cdots U(m_j))}$ is by noting that the 
linear coordinates $a_i$ of $\s {A}$  in (\ref{asing})
correspond to variations $\delta v_i$ of the local coordinates 
$v_i$ of $\S {S(U(m_1) \times \cdots U(m_j))}$ in equation (\ref{dfadj}).
Theorem~I.e is therefore verified in this case.\\
We  give more details for the special cases $N=3$ and $N=4$.\\

\noindent
$SU(3)$ {\em with an adjoint field:} The partitions 
of $N=3$ are completely ordered:
$$ 3 > 2+1 > 1+1+1$$
Equivalently, we have the following ordered set of isotropy groups: 
$$SU(3) > U(2) > U(1) \times U(1)$$
leading to the arrangement  $$\S {U(1) \times U(1)} - \S {U(2)} - \S {SU(3)}$$
of the strata, which have complex dimensions $2,1$ and $0$.
The equations defining the strata of $\mc \simeq \c^2$ can be obtained 
by finding the relations among the invariants $t_j$ at  points $A_H$
of the form
(\ref{dfadj}-\ref{dfadj1}) with  isotropy group $H$:
\begin{eqnarray}
\left(A^{\a}_{\b}\right)_{SU(3)} &=& 0; 
\;\;\; 
\left(A^{\a}_{\b}\right)_{U(2)} = 
\text{diag} (x,x,-2x), \; x \neq 0; \\ \nonumber  
\left(A^{\a}_{\b}\right)_{U(1) \times U(1)} &=& \text{diag} (x,y,-x-y), 
y \neq x, -2x, -x/2. 
\end{eqnarray}
 For example, at
 $\left(A^{\a}_{\b}\right)_{U(2)}$ we have 
$t_1=6x^2, t_2=-6x^3, x \neq 0$, this defines the algebraic 
set $t_1^3-6t_2^2=0$ with the point $(0,0)$ removed.
Proceeding in  this way we arrive at
\begin{eqnarray}
\S {U(1) \times U(1)} &=&\{(t_1,t_2) \in \c^2 | 
t_1^3-6t_2^2 \neq 0 \}, \\ \nonumber
 \S {U(2)} &=& \{(t_1,t_2) \in \c^2 | 
t_1^3-6t_2^2 = 0 \; \text{and} \; (t_1,t_2) \neq (0,0) \}, \\ \nonumber
\S {SU(3)} &=& \{(0,0)\}. 
\end{eqnarray}
\vspace{.5cm}

\noindent
$SU(4)$ {\em with an adjoint:} we have the following partitions of $4$:
\begin{equation} \begin{array}{ccccccc}
& & 3+1 & & &&\\
 & \diagup & & \diagdown &  &  &  \\
4 &  &  &  & 2+1+1 & \; \; - \;\; & 1+1+1+1, \\
   & \diagdown  && \diagup &&&\\
&& 2+2 &&&& \end{array} \end{equation}
corresponding to the following patterns of symmetry breaking
\begin{equation} \begin{array}{ccccccc}
& & U(3) & & &&\\
 & \diagup & & \diagdown &  &  &  \\
SU(4) &  &  &  & U(2) \times U(1) & \;\; - \;\;
 & U(1) \times U(1) \times U(1). \\
   & \diagdown  && \diagup &&&\\
&& S(U(2) \times U(2)) &&&& \end{array} \end{equation}
Following branches from left to right be have two decreasing 
sequences of isotropy groups, or two increasing sequence of strata
of dimensions $0,1,2$ and $3$.
There is no order relation between the one dimensional
$U(3)$ and 
$S(U(2) \times U(2))$ strata.
 Generic diagonal elements at  different strata have the forms 
\begin{eqnarray}  \label{stru4}
&&\left(A^{\a}_{\b}\right)_{SU(4)} = 0; \\ \nonumber
&& \left(A^{\a}_{\b}\right)_{U(3)} =
\text{diag} (x,x,x,-3x),\; x \neq 0; \\ \nonumber  
&&\left( A^{\a}_{\b}\right)_{S(U(2) \times U(2))} = \text{diag} (x,x,-x,-x),
\; x \neq 0;  \\ \nonumber
&& \left(A^{\a}_{\b}\right)_{U(2) \times U(1)} = \text{diag} (x,x,y,-2x-y),
\; y \neq \pm x, -3x; \\ \nonumber
&&\left(A^{\a}_{\b}\right)_{U(1) \times U(1) \times U(1)} = \text{diag}
(x,y,z,-x-y-z), x,y,z \;\; \text{and} \;\; -x-y-z \;\; \text{all different}.
\end{eqnarray}
From the above equations we get $t_1=2x^2+y^2+(2x+y)^2,
t_2=2x^3+y^3-(2x+y)^3,$ and $t_3=2x^4+y^4+(2x+y)^4$ at
$\S {U(2) \times U(1)}$. If $x$ and $y$ are unrestricted,
these are parametric equations
for $\overline{\S {U(2) \times U(1)}} \subset \c^3$. An
equivalent implicit equation, obtained by using Gr\"oebner basis~\cite{clo},
is $288t_3t_1^2+144t_3t_1t_2^2
-90t_3t_1^4-288t_3^3+9t_1^6-68t_2^2t_1^3-24t_2^4=0$.
The equations defining the strata are
\begin{eqnarray}       \nonumber
\S {U(1) \times U(1) \times U(1)} &=& \{(t_1,t_2,t_3)  |
288t_3t_1^2+144t_3t_1t_2^2
-90t_3t_1^4-288t_3^3+9t_1^6-68t_2^2t_1^3-24t_2^4 \neq 0 \} \\ \nonumber
\S {U(2) \times U(1)} &=& \{(t_1,t_2,t_3) |
288t_3t_1^2+144t_3t_1t_2^2
-90t_3t_1^4-288t_3^3+9t_1^6-68t_2^2t_1^3-24t_2^4 = 0 \\ \nonumber &&\;\;
\text{ and } ( t_2 \neq 0 \; \text{ or }
t_1^2-4t_3 \neq 0 ) \text { and }
( t_1^3-3t_2^2 \neq 0 \text{ or }  \frac{7}{12}t_1^2-t_3 \neq  0 )
\}  \\ \nonumber
\S {S(U(2) \times U(2))} &=&\{(t_1,t_2,t_3)  |
t_2 = 0, t_1^2-4t_3=0,  \text { and } t_3 \neq 0  \}, \\ \nonumber
 \S {U(3)} &=& \{(t_1,t_2,t_3)  |
t_1^3-3t_2^2 = 0,  \frac{7}{12}t_1^2-t_3 = 0, \; \text{ and }
t_3 \neq 0  \}, \\ \nonumber
\S {SU(4)} &=& \{(0,0,0)\}. 
\end{eqnarray}
$\overline{\S {U(2) \times U(1)}}$ is a two dimensional complex surface
on which the complex curves $\overline{\S {U(3)}}$ and $\overline{\S
{S(U(2) \times U(1)}}$ lie. These two curves meet at $\S {SU(4)}$.  \\

\noindent
Our final example is a theory with an unstable representation 
of the complexified gauge group. \\
\noindent
{\it Example \ref{exs}.4:} Let $G=SU(2N+1), \rho =  \overline{\Yasymm} 
+ (2N-3) \overline{\Yfund}$, the classical flavor symmetry group is
$K = U(1) \times U(2N-3)$. If $N=2$, the only \df point is
the trivial one, and  $\mc$ is a zero dimensional vector space. Actually,
the $SU(5)$ with an antifundamental and an antisymmetric tensor, 
together with $SO(10)$ with a spinor, are the only theories based 
on a simple gauge group with only trivial \df points, and therefore 
a single stratum. If $N \geq 3$, $\mc$  
 is the vector space  of  $U(2N-3)$
unconstrained antisymmetric tensors 
$V^{ij} = A^{\a \b} Q^i_{\a} Q^j_{\b} = \pi(Q,A)$. The $D-$flatness condition 
reads tr~$ [T(2A A^{\dagger} - Q^{\dagger} Q)] = 0$. 
A generic \df point can be $G \times K$
rotated to 
\begin{eqnarray} \label{dfunst} 
Q^i_{\a} &=& \left( \begin{array}{cc} q & 0 \\ 0 & 0 \end{array} \right), 
\;\;\; q = diag(q_1,q_2,...,q_{2k}), \\ \nonumber
A^{\a \b} &=& \left( \begin{array}{cc} v & 0 \\0&0 \end{array}
 \right), \;\; v = \text{diag}(v_1 \sigma, v_2 \sigma, ..., v_k \sigma), \;\; 
 k \leq N-2, \;\; 
\sigma = \left( \begin{array}{rr} 0 & 1 \\ -1 & 0 \end{array} 
\right), 
\end{eqnarray}
with $|q_{2j-1}| = |q_{2j}| = |v_j| \neq 0$. This point breaks $G$ to
$SU(2(N-k)+1)$, the set of  strata
$\S {SU(2(N-k)+1)}, k=0,...,N-1$ being  totally ordered.
Under $\pi$, (\ref{dfunst}) goes to 
\begin{equation} \label{p}
V^{ij} = \text{diag} ( q_1q_2v_1 \sigma, q_3q_4v_2 \sigma,...,q_{2k-1}q_{2k} 
v_k \sigma,0,0,...,0). \end{equation}
The $K$ orbits of the points (\ref{p}) generate the $SU(2(N-k)+1)$ stratum.
$\S {SU(2(N-k)+1)}$ is 
 the  $4kN-2k^2-7k$ dimensional  
complex manifold of $(2N-3) \times (2N-3)$ antisymmetric
matrices $V^{ij}$ of rank $2k$.\\
 Under $SU(2(N-k)+1)$, the configuration
space $\c ^{(2N+1)(3N - 3)} \simeq T_{(A,Q)} \c ^{(2N+1)(3N - 3)} $
breaks into $\t {(A,Q)}
\oplus \s {(A,Q)}
\oplus \n {(A,Q)}$. Using   (\ref{dfunst})
and writing a  $\lie {G^c}$ element as
\begin{equation}
T = \left( \begin{array}{cc} t_1 & t_2 \\ t_3 & t_4 \end{array} \right), 
\hspace{1cm} t_4 \in \lie {SL(2(N-k)+1)}, 
\end{equation}
we obtain
$$ \t {(A,Q)}:
\delta Q = \left( \begin{array}{cc} -q t_1 & -q t_2 \\ 0 & 0 \end{array}
\right), \hspace{1cm} \delta A = \left( 
\begin{array}{cc} t_1v + vt_1^T &
v t_3^T \\ t_3 v & 0 \end{array} \right). $$
A possible choice for $\n {(A,Q)} \oplus \s {(A,Q)}$ is
\begin{eqnarray} \label{sliceunst}
\s {(A,Q)}: \;\;\;  \delta Q = \left( \begin{array}{cc} 0&0 \\
\delta q_3 & 0 \end{array} \right), 
\hspace{1cm} \delta A = \left( \begin{array}{cc} \delta A_1 & 
0 \\ 0 & 0 \end{array} \right) \\ \nonumber
\n {(A,Q)}: \;\;\; 
         \delta Q = \left( \begin{array}{cc} 0 & 0 \\ 0 & 
\delta q_4 \end{array} \right), \hspace{1cm} 
\delta A = \left( \begin{array}{cc} 0 & 
0 \\ 0 & \delta A_4 \end{array} \right)
\end{eqnarray}
The special feature of this example is that the $G^c$ action 
is {\em unstable}. Although  $G^c$ applied to  
(\ref{dfunst}) with $k=N-1$ gives  a highest dimensional
$G^c$ orbit containing \df points,
there are  $G^c$ orbits of higher dimension.
An example of a highest dimensional  orbit is that
 of the 
configuration point 
\begin{eqnarray} \label{nco}
Q^i_{\a} &=& \left( \begin{array}{ccc} {\bf 0}_{(2N-3) \times 3}&
q & {\bf 0}_{(2N-3) \times 1}   
\end{array} \right), 
\;\;\; q = diag(q_1,q_2,...,q_{2N-3}), \\ \nonumber 
A^{\a \b} &=& \text{diag}(v_1 \sigma, v_2 \sigma, ..., v_N \sigma, 0),  
\hspace{1cm} 
\sigma = \left( \begin{array}{rr} 0 & 1 \\ -1 & 0 \end{array} 
\right).
\end{eqnarray}
The $G^c$ isotropy group at (\ref{nco}), ${G^c}_0$, 
is different from ${G_0}^c$, a common situation
for the $G$ and $G^c$ isotropy groups at $G^c$ orbits of non
\df points.
 We can readily check that $\lie {{G^c}_0}$ is
the set of $T \in {\frak sl}(2N+1,\c)$ having  the form
\begin{equation} \label{nrunst}
T = \left( \begin{array}{rrrrrrrr} x & y & 0 & a & 0 &
 \cdots & 0 & d \\ z & -x & 0 & b&0& \cdots & 0&e\\ \frac{-v_2}{v_1}b
&\frac{v_2}{v_1}a&0&c&0&\cdots&0&f \\
0&0&0&0&0&\cdots&0&0 \\
\vdots & \vdots & \vdots & \vdots & \vdots & & \vdots& \vdots \\
0&0&0&0&0&\cdots&0&0 \end{array} \right). 
\end{equation}
$x,y$ and $z$ span an ${\frak sl}(2,\c)$ non-invariant Lie subalgebra of
the isotropy subalgebra, whereas $a,b,c,d,e,f$ 
span a six dimensional unipotent (a Lie algebra of nilpotent matrices) 
Lie algebra ${\frak u}_6$ which is an ideal of $\lie {{G^c}_0}$. 
In other words
\begin{equation} 
\lie{{G^c}_0} = {\frak sl}(2,\c) \oplus {\frak u}_6 \; \text{(direct sum
of vector spaces)}, \;\; [\lie{{G^c}_0}, {\frak{u}}_6] \subseteq {\frak u}_6.
\end{equation}
After exponentiating we get a semidirect product: $G^c_0 = 
SL(2,\c) \ltimes {\frak U}_6$.  \\
The slice representation (\ref{sliceunst}) at the \df point
eq.(\ref{dfunst})
is $SU(2(N-k)+1)$ with $[2(N-k)-3] \overline{\Yfund} + \Yasymm 
+ (4kN-2k^2-7k) {\Bbb I}$. At the main stratum, $k=N-2$,
the slice is $SU(5)$ with $ \overline{\Yfund} + \Yasymm + (2N-3)(N-2) 
{\Bbb I}$. Taking out the singlets we get
$SU(5)$ with $ \overline{\Yfund} + \Yasymm$, a theory
with a zero dimensional moduli space,
theorem~I.f is verified.
To show that $SU(5)$ with $ \overline{\Yfund} + \Yasymm$
has a zero dimensional moduli space we specialize
the above equations to the $N=2$ case. The orbit
of (\ref{nco}) has dimension $15$, as its isotropy group
(\ref{nrunst}) has dimension $9$.
Taking the closure of this orbit we a get a fifteen
dimensional algebraic subset of $\c^{15} \simeq  \overline{\Yfund} + \Yasymm$,
the only possibility being the whole
$ \overline{\Yfund} + \Yasymm = \{ \phi \}$
vector space. If $\hp(\phi)$ is a holomorphic invariant, then $\hp(\phi)$
is constant in the closure of this orbit, i.e., the only holomorphic invariants
of this theory are the constants, $\mc$ is a zero dimensional vector space.

\section{Applications} \label{app}
\subsection{Low energy construction of $\mc^W$ and Lagrange multipliers}
\label{lec}
A holomorphic $G$ invariant superpotential $W: \c^n \to \c$
can always be written in terms  of a basic set of holomorphic  invariants
$\hp^i(\phi), i=1,...,s$, as $W(\phi) = \hw (\hp(\phi))$, $\hw$ being an
arbitrary $\c^s \to \c$ function.
The $W=0$ classical moduli space $\mc$ is parameterized by the subset of $\c^s$
defined by the algebraic constraints $p_{\a}(\hp)=0, \a=1,...,l$ among the
basic invariants $\hp(\phi)$.
The moduli space $\mc^W$ of
the supersymmetric gauge theory with the added superpotential is
usually obtained by first solving for  the $F$-flat point  set $\c^n_W =
\{ \phi \in \c^n | dW(\phi)=0 \}$, then projecting $\c^n_W$
 down to $\c^s$ using the map $\pi: \phi \to \hp(\phi)$, i.e.,
 $\mc^W = \pi(\c^n_W)$.
It can be shown~\cite{luty} that  $\mc^W \subset \mc \subseteq \c^s$ is the
the algebraic set defined by
the gauge invariant polynomial constraints
$p_{\a}(\hp)=0, \a=1,...,l; w_{\b}(\hp)=0,
\b=1,...,r$,
where $w_{\b}(\hp)=0, \b=1,...,r$ are the gauge invariant
constraints resulting from $dW=0$~\cite{luty}.
In this section we elaborate further on  the results in~\cite{plb}
on  methods to obtain from $\hw$ and $p_{\a}(\hp)=0$ the equations
$w_{\b}(\hp)=0$ defining
${\cal M}^W \subset \mc \subseteq \c^s$, working
entirely in the space $\c^s$ of composite
superfields $\hp$, assuming
we do not know the functions  $\hp(\phi)$, i.e., how 
the composite superfields are made out of the elementary fields.
In Section~\ref{intro} we used an $SO(N)$ theory with two $\Yfund$
to show that  knowledge of $\hw$ and the constraints among
the basic invariants is not enough to obtain $\mc^W$,
and  claimed that
the required additional information was the stratification
of the moduli space. This last assertion follows from Theorem~I:
the differential at the \df point $\phi$ of the map $\pi: \phi \to \hp(\phi)$,
$\pi'_{\phi} = \partial \hp^j(\phi) / \partial \phi^i$,
annihilates the subspace $\t {\phi} \oplus \n {\phi}$
of $\c^n = \t {\phi} \oplus \n {\phi} \oplus \s {\phi}$,
(Theorem~I.e) and so
\begin{equation} \label{w'}
\frac{\partial W}{\partial \phi^i} \delta \phi^i =
\frac{\partial \hw}{\partial \hp ^j} \left( \frac{\partial \hp ^j}{\partial
\phi^i} \delta \phi^i \right), \;\;\;\;
\left( \frac{\partial \hp ^j}{\partial \phi^i} = \pi' \right),
\end{equation}
is zero if $\delta \phi \in \t {\phi} \oplus \n {\phi}$.
On the other hand, again by Theorem~I.e,
$\partial \hp^i(\phi) / \partial \phi^j \, \delta \phi^j$
does not span the whole tangent space $T_{\hp(\phi)}\mc$
of $\mc$ at $\hp(\phi)$,
but only the subspace $T_{\hp(\phi)} \S {(G_{\phi})} \subseteq T_{\hp(\phi)}
\mc$ tangent to the stratum
through $\hp(\phi)$.
Therefore, $dW=0$ is equivalent to
\begin{equation} \label{dw}
 \frac{\partial \hw}{\partial \hp^i} {\Bigg\vert_{\hp(\phi)}} \delta \hp^i
= 0, \;\;\; \forall \delta \hp^i \in T_{\hp(\phi)} \S {G_{\phi}},
\end{equation}
In other words,  $dW(\phi)=0$ if and only if $\pi(\phi)$
is a stationary point of the restriction $\hw_{(G_{\phi})}
\equiv \hw|_{\S {(G_{\phi})}}$ of $\hw$ to
the stratum passing through $\pi(\phi)$.
This fact, pointed out  in~\cite{plb} gives an answer to the problem
of finding $\mc^W$ working entirely with gauge invariant operators:
first  find, for each stratum $\S {(H)}$,  the critical points
of the restriction of $\hw$ to $\S {(H)}$,
then take the union of the resulting sets. We will see in the
following section that it is not always necessary to solve the
stationary point
equations at {\em every} stratum.
There are two ways of finding the stationary points of
 $\hw_{(H)} \equiv \hw_{|_{\S {(H)}}}$.
We can use the fact that $\S {(H)}$ is a complex manifold, cover
it with local coordinate charts $\{ x^i, i=1,...,\text{dim } \S {(H)} \}$,
and find the critical points $\partial W_{(H)} / \partial x^i=0$ in every chart.
Alternatively, we can use  Lagrange multipliers and find the critical points
of $\hw_{(H)} + c^{\beta} K^{(H)}_{\beta}$.
Here $K^{(H)}_{\beta}(\hp)=0$ are the
 equations (partially)
defining $\S {(H)}$. In fact the $K^{(H)}_{\beta}(\hp)$
are  polynomials, their zero set is
the smallest algebraic set containing $S {(H)}$, i.e., the
Zariski closure $\overline{\S {(H)}}$ which, according
to Theorem~I.b, is the union of $\S {(H)}$ and the
smaller dimensional strata in its boundary. Any stationary point of
$\hw_{(H)} + c^{\beta} K^{(H)}_{\beta}$ outside $\S {(H)}$ has
to be discarded.
The  Lagrange multiplier method is ``safe"   because
it only  requires  that the constraints $K_{\b}^{(H)}(\hp)$
satisfy the condition
rank~$\partial K^{(H)}_{\b} / \partial \hp ^j
=$~maximal.
As $\S {(H)}$ is a complex manifold,
points in $\S {(H)} \subseteq \overline{\S {(H)}}$
are smooth, and the rank condition is met at the stationary points that
are not discarded.
This guarantees the validity of applying
Lagrange multipliers to this problem.\\

\noindent
{\it Example \ref{lec}.1:}
Assume a given theory  contains no $G$ singlets, then $\S {(G)} = \{\hp=0 \}$
is zero dimensional and $d \hw_{|_{\S {(G)}}}=0$ is trivially
satisfied, thus   $\S {(G)} \subseteq \mc^W$.
In a  microscopic description  we prove $0 = \hp(0) \in \mc^W$
by noting that, since there are no gauge  singlets,
  $\hp(\phi)$ is at least quadratic in $\phi$ and
so $dW$ eq.~(\ref{dw}) equals zero  at the \df point $\phi=0$.\\

\noindent
{\it Example \ref{lec}.2:} Consider  the  $SO(N)$ with $2\Yfund$ theory.
The basic invariants are 
 $S_{ij} = Q^{\a}_{i} Q^{\a}_{j}$, $\mc = \{S_{ij} \} = \c^3$. 
There are three strata: 
\begin{eqnarray} \label{sostrat}
\S {SO(N-2)} &=& \{ S | \;\text{det}\; S \neq 0 \}, \\ \nonumber 
\S {SO(N-1)} &=& \{ S \neq 0 |\; \text{det}\; S \; = 0 \}, \\ \nonumber
\S {SO(N)} &=& \{ S = 0 \}. 
\end{eqnarray} 
The polynomials $K^{(H)}_{\b}$ in  the definition of the strata are
$K_1^{SO(N-1)}= S_{11}S_{22} - {S_{12}}^2; \;\;
K_{ij}^{SO(N)}= S_{ij}, (i,j)=(1,1),(1,2),(2,2)$, no constraints for
$\S {SO(N-2)}$. The equation $\text{det } S = 0$
actually defines the
{\em closure} of  $\S {SO(N-1)}$ where
 $\partial (\text{det } S)/ \partial S_{ij}$ fails to have constant rank
because of the included   boundary point $S=0$. The additional condition
$S \neq 0$ in the definition of $\S {SO(N-1)}$
 excludes the boundary, problematic point
 that would
invalidate the Lagrange multipliers method.\\
Assume $\hw(S_{ij}) = mS_{22}$. We will find $\mc^W$ using the
two methods described above.
(i) \underline{Local charts on the strata:}  \\
{\em Vacua at $\S {SO(N-2)}$:}
$\S {SO(N-2)}$ is an open subset of $\c^3$, $\{(S_{11},S_{12},S_{22})\}$
is an appropriate set of (global) coordinates. There are no critical
points  of $\hw_{SO(N-2)}(S_{11},S_{12},S_{22}) = mS_{22}$,
there is no vacuum at the principal stratum. \\
{\em Vacua at $\S {SO(N-1)}$:} $\S {SO(N-1)} $ can be
covered with two coordinate patches: 
$\S {SO(N-1)}^{(A)}$, the set defined by $S_{11} \neq 0$ and
$\S {SO(N-1)}^{(B)}$, the open subset where
$S_{22} \neq 0$. The  coordinates are
\begin{equation}
S_{ij} = \left( \begin{array}{cc} x & y \\ y & y^2/x \end{array} \right) 
\;\;x \neq 0 \; \text{on } \S {SO(N-1)}^{(A)}, \;
\;\;
S_{ij} = \left( \begin{array}{cc} y^2/z & y \\ y & z \end{array} \right) 
\;\; z \neq 0 \; \text{on } \S {SO(N-1)}^{(B)}.
\end{equation}
We find  that ${\hw} _{SO(N-1)}(y,z) = mz$ at the $B$ chart,
 $d {\hw} _{SO(N-1)} = 0$ has no solutions there.
At $\S {SO(N-1)}^{(A)}$, $\hw _{SO(N-1)}(x,y) = my^2/x$, and
we find the solutions $S_{ij}= \text{diag} (x,0), x \neq 0$.\\
{\em Vacua at $\S {SO(N)}$:} the only point of this zero dimensional
manifold is a vacuum.\\
Taking the union of the solution sets we arrive at:
\begin{equation} \label{mod}
\mc ^W = \{ S_{ij} | S_{12} = S_{22} = 0 \}.
\end{equation}
(ii) \underline{Lagrange multipliers:} \\
{\em Vacua at $\S {SO(N-2)}$:} we find the extrema of  $f(S_{11},S_{12},
S_{22}) = m S_{22}$ and keep only
the solutions satisfying det~$S \neq 0$. There are no solutions. \\
{\em Vacua at $\S {SO(N-1)}$:} we  find the extrema of $
f(S_{11},S_{12},S_{22}) = m S_{22} + \alpha
(S_{11} S_{22} - S_{12}^2)$ and discard $S=0$ as a solution.
The solutions  are
$\a \neq 0, S_{ij} = \text{diag}\; (-m/\a,0)$. \\
{\em Vacua at $\S {SO(N)}$:} we look for stationary points of 
 $f(S_{11},S_{12},S_{22}) =
mS_{22}+\a S_{11} + \b S_{12} + \gamma S_{22}$ and find
$S_{ij}=\a=\b=m+\gamma=0$.\\
Taking the union of the solution sets we recover~(\ref{mod}).

\subsection{Irreducible components of $W \neq 0$ moduli spaces}
\label{comp}

An algebraic set is said
to be irreducible if it is not  the union of
two distinct algebraic sets. Every algebraic set $X$ can be
uniquely
decomposed as $X = \cup_{i=1}^r X_i$, with   $X_i$ irreducible
and $r$ minimal.
As an example, the set $X \subset \c^2 = \{(x,y) \}$ defined
by the equation $xy=0$ has two irreducible components: $X = \{(x,y)| x=0\}
\cup \{ (x,y)| y=0 \}$.  The moduli space $\mc$ of a supersymmetric
gauge theory with zero superpotential is irreducible, because is the
image under the regular (polynomial) map $\pi$ of the irreducible
set $\c^n$~\cite{clo}, the vector space of elementary fields.
However, when a superpotential is added, $\mc^W$ is generically
reducible. We will see that complete irreducible components of
$\mc^W$ can be obtained by finding  their vacua just at the maximal stratum
intersecting the component,
instead of searching in every stratum. This is particularly
useful if $\mc^W$ is known a priori to be irreducible, case in which
we will only need to solve the equation $d \hw _{(H)}=0$ in a
single stratum.
A trivial example
of an irreducible  moduli space $\mc^W$ is when $\mc^W$
consists a single point. Such theories are interesting because they may lead to
dynamical supersymmetry breaking in the quantum regime~\cite{dsb}.
Another example arises in the process of
 integrating out heavy composites from an
effective superpotential $W_{eff}$. A tree level mass term $W_{mass}=m\hp^1$
is added to a supersymmetric gauge theory whose low energy effective
superpotential $W_{eff}(\hp)$ is known. The effective superpotential
of the resulting theory is obtained by integrating out the heavy
composites $\hp^i, i=1,2,...,r \leq s$ from $W_{eff}$, usually
identified from the elementary field content of $\hp^1$ and the other
invariants. The heavy composites can also
 be identified using the stratification
of the zero superpotential classical moduli space
$\mc = \{ \hp \in \c^s | p_{\a}(\hp)=0 \}$,
without knowing the elementary quark content of the invariants.
The light elementary fields $\phi$ span the  vector space $\c^n_{W_{mass}} =
\{ \phi \in \c^n | \partial W_{mass}
/ \partial \phi^i = 0\}$, which is irreducible,
then $\mc^{W_{mass}}= \pi(\c^n_{W_{mass}})=\{ \hp \in \c^s | p_{\a}(\hp)=0,
\; \text{ and } \; \hp^j=0, j=1,...,r \}$ is also irreducible, and excludes
precisely the heavy fields to integrate out from $W_{eff}$.
The problem of identifying heavy composites
reduces to finding the irreducible classical moduli space
$\mc^{W_{mass}}$, which can be done using the stratification of $\mc$.
For irreducible moduli spaces $\mc^W$,
important simplification arise in the  methods described in~\cite{plb}.\\
Let
\begin{equation} \mc^W = \bigcup_i {\mc^W}_{(i)} \end{equation}
be the decomposition  of $\mc^W$ into irreducible components. As
proved in Appendix~A, the set  of strata intersecting
${\mc^W}_{(i)}$ contains a unique maximal element $\S {(H_i)}$.
Furthermore (eq~(\ref{muf}))
\begin{equation}       \label{icms}
{\mc^W}_{(i)} = \overline{{\mc^W}_{(i)} \cap \S {(H_i)}}.
\end{equation}
The above equation tells us that once the maximal set intersecting
${\mc^W}_{(i)}$ is found, we only need to  find the stationary points of
$\hw_{(H_i)}$ and take the closure of the resulting set. In taking
the closure, we are actually incorporating all the other vacua in
the smaller strata intersecting ${\mc^W}_{(i)}$
without solving the corresponding stationary point
equations. If $\mc^W$ is irreducible, we only need to solve the equation
$d \hw _ {(H)} = 0$  on a single stratum (the maximal stratum intersecting
$\mc^W$),
then take the closure of the critical point set,  otherwise we follow the
procedure described below.

\subsubsection{Procedure to obtain $\mc^W$}

This procedure  is based on the fact that the set of strata
intersecting an irreducible component ${\mc^W}_{(i)}$ of the
moduli space contains a single maximal element $\S {(H_i)}$
and eq~(\ref{icms}) holds. It stops after a few steps
if $\mc^W$ is irreducible.\\

\noindent
{\bf Procedure to obtain ${\mathbf \mc^W}$:}
$\mc^W \subset \mc \subseteq \c^s$ can be
 obtained, one (subset of)
irreducible component(s) at a time, by means of  the following
procedure:\\
\begin{itemize}
\item [[i]] Arrange  the partially ordered set of strata of $\mc$
as explained at the beginning of Section~\ref{exs}.
 By Theorem~I.c
the first and last columns contain a single entry ($\S {(G_P)}$ and
$\S {(G)}$ respectively). The set of paths through  linked strata
give all the  different patterns of gradual  symmetry breaking from $G$ to
$G_P$.
\item [[ii]] Look for solutions of $d {\hw}_{(G_P)} = 0$.
If there are solutions, take the closure of the solution
set $\{ \hp \in \S {(G_P)} | d {\hw}_{(G_P)}(\hp) = 0 \}$,
this yields one or more complete  irreducible components of $\mc^W$.
\item [[iii]] Look for new solutions in the strata in the next column,
if there are new solutions, say in $\S {(H)}$,
go to [iv], otherwise repeat [iii].
\item [[iv]] Take the closure of the solution set
to obtain further irreducible components of  $ \mc ^W$.
\item [[v]] Look for new solutions in the other strata in the column
of $(H)$, if any, go to [iv],
otherwise go to [iii]
\end{itemize}

Solutions to $d \hw _{(H)} =0$ can be found either by covering
the stratum with local coordinates or by using Lagrange 
multipliers, as explained above.
Step iv saves us some work, in taking the closure we obtain some solutions 
 $dW_{(H')}=0, (H') > (H)$ without actually performing explicit
computations. However,
if $\mc^W$ is reducible,
$\overline{\mc^W \cap \S {(H)}}$ does not
necessarily
exhaust the solution set $\bigcup_{(H') \geq (H)} (\mc^W \cap \S {(H')})$.
The following example exhibits some of these subtleties.\\

\noindent
{\it Example \ref{comp}.1:} {\em $SO(13)$ with a spinor (Figure 1):}
A complete classification of the $G^c$ orbits  of this theory
can be found in ref~\cite{gv}. Theorem~I
  in \cite{gv} states  that
there are two invariants, $p$ and $q$ (of degrees $4$ and $8$ 
in the elementary spinor) which are unconstrained, i.e., 
$\mc = \c^2$. There are four strata (as there are four
types of closed $G^c$ orbits, the ones that contain \df points,
see Table~1 in \cite{gv}),
we  order them as
in step [i] of the procedure above:
\begin{equation} \label{sspin13} \begin{array}{cccccc}
& & \S {G_2 \times SU(3)} & & &\\
 & \diagup & & \diagdown &  &    \\
\S {SU(3) \times SU(3)}&  &  &  && \S {SO(13)}. \\
   & \diagdown  && \diagup &&\\
&& \S {SU(6)} &&& \end{array} \end{equation}
The equations defining the strata are the following 
\begin{eqnarray}   \label{ssspin13}
\S {SU(3) \times SU(3)} &=& 
\{(p,q) | p^2-4q \neq 0 \; \text{and} \; q \neq 0 \} \\ \nonumber
\S {G_2 \times SU(3)} &=& 
\{(p,q) | p^2-4q = 0 \; \text{and} \; p \neq 0 \} \\ \nonumber
\S {SU(6)} &=& 
\{(p,q) | q = 0 \; \text{and} \; p \neq 0 \} \\ \nonumber
\S {SO(13)} &=& \{(0,0)\}.
\end{eqnarray}
The real section $(p,q) \in {\Bbb R}^2$ of $\mc \simeq \c^2$ and
its strata is
depicted in Figure~1.a.
The dimensions of the strata in the first, second and third column of
(\ref{sspin13}) are
respectively two, one  and zero.
We will not use Lagrange multipliers but local coordinates on the strata.
$\{ (p,q) | q \neq 0, p^2/4 \}$ is a good set of
 (global) coordinates on the principal stratum,
whereas $p \neq 0$ can be taken as a (global)
coordinate of $\S {SU(6)}$ and also
of $\S {G_2 \times SU(3)}$.
 We apply the procedure above to solve for
$\mc ^W$ in the following three cases (step {[i]} is already done in
equation~(\ref{sspin13})):\\

\noindent
{\em (i) $ \hw (p,q) = f(p)$ (Figure 1.b).} \\
{\bf step {[ii]}:} $\hw_{SU(3) \times SU(3)}(p,q) = f(p), q \neq 0, p^2/4$.
 The set of critical points is
 $\mc^W \cap \S {SU(3) \times SU(3)} = \{ (p_i,q) |
q \neq 0, p_i^2/4 \text{  and  } f'(p_i)=0, i=1,...,k \}$, $k$
the number of distinct roots of the polynomial $f'$.
The closure of this set is $\{(p_i,q)  | q \in \c, i=1,...,k \}$, which
is the union of $k$ irreducible sets. \\
{\bf step {[iii]}:} No {\em new} solution arises in $\S {G_2 \times SU(3)}$ or
$\S {SU(6)}$  but those already found in taking the closure in step {[ii]}. \\
{\bf step {[iii]}:} If $0$ is among the $p_i$'s, there is not any new solution
in $\S {SO(13)}$,
otherwise we add the solution $(p,q)=(0,0)$.\\
\begin{equation} \label{dec-i}
\mc ^W = \cup_{i=1}^k \{ (p_i,q) | q \in \c \} \cup \{(0,0)\},
\end{equation}
has  $k+1$ irreducible components if $f'(0) \neq 0$, $k$ components if
$f'(0)=0$. \\

\noindent
{\em (ii) $ \hw (p,q) = (p^2-4q-m^8)^2/M^{13}$, $m\neq 0$ (Figure 1.c).}\\
{\bf step {[ii]}:} $\hw _{SU(3) \times SU(3)} = \hw(p,q)$ with the restrictions
$q \neq 0, p^2/4$,  $d \hw _{SU(3) \times SU(3)} = 0$ gives
$\mc ^W \cap \S {SU(3) \times SU(3)} = \{(p,q) | q = (p^2-m^8)/4)
\text{ and }  p \neq \pm m^4 \}$. The closure of this set is
$\{ (p,q) | q = (p^2-m^8)/4 \}$. \\
{\bf step {[iii]}:} $\hw_{SU(6)} (p) = (p^2-m^8)^2/M^{13}$, $p \neq 0$.
$d \hw_{SU(6)} = 0$ only at 
$p=\pm m^4$.
These two solutions correspond  to
$\overline{\left(\mc ^W \cap \S {SU(3) \times SU(3)}\right) } 
\cap \S {SU(6)}$, they are not {\em new} solutions,
 we are still
seeing  the irreducible component of $\mc^W$ found in step {[ii]}.
Contrast with what happens at
$\S {G_2 \times SU(3)}$.  $\hw_{G_2 \times SU(3)} = m^{16}/M^{13} =
constant$, then $d \hw_{G_2 \times SU(3)} \equiv 0$. 
$\S {G_2 \times SU(3)} \subset \mc^W$ is  an entire 
new set of solutions!  
In fact  $\overline{\mc ^W \cap \S {SU(3) \times SU(3)}} \cap \S 
{G_2 \times SU(3)} = \emptyset$. \\
{\bf step {[iv]}:} In taking the closure of $\mc^W \cap \S {G_2 \times SU(3)}$
we add the solution $(0,0)$ that completes the $q=p^2/4$ parabola.\\
{\bf step {[v]}:}
 We go back to step {[iii]} and find the trivial solution at $\S {SO(13)}$,
which is not new.\\
$\mc ^W$ has 
two irreducible components:
\begin{equation} \label{dec-ii}
 {\mc ^W}_{(1)} = \{ (p,q) : q = (p^2-m^8)/4 \}, 
\;\;\;  {\mc^W}_{(2)} =  \{ (p,q) | q = p^2/4 \}.
\end{equation}

\noindent
{\em (iii) $\hw (p,q) = [p(p-\a)-q]^2/M^{13}$ (Figure 1.d)}. \\
This example is somewhat 
intermediate between {\em (i)} and {\em (ii)} in the sense that 
 the closure of the solution 
set in a given stratum intersects smaller strata, where also new solutions
 arise. The superpotentials and solution sets at different strata are:
\begin{eqnarray} \begin{array}{ll}   \nonumber
\hw _{SU(3) \times SU(3)} = \frac{[p(p-\a)-q]^2}{M^{13}},  & \mc^W \cap \S {SU(3) \times
SU(3)} = \{ (p,q) | q=p(p-\a)), \;  q \neq 0,p^2/4 \}; \\  \nonumber
\hw _{SU(6)} =  \frac{[p(p-\a)]^2}{M^{13}}, &  \mc^W \cap \S {SU(6)}
 = \{ (\a,0), (\a /2,0) \};  \\           \nonumber
\hw _{G_2 \times SU(3)} = \frac{[\frac{3}{4}p^2-p \a]^2}{M^{13}}, &
\mc^W \cap \S {G_2 \times SU(3)} = \{ (2 \a /3,\a ^2 /9), 
(4 \a /3, 4 \a^2 /9) \};  \\ \nonumber
\hw _{SO(13)} = 0, &  \mc \cap \S {SO(13)} = \{ (0,0) \}. \end{array} 
 \label{iii} \end{eqnarray} 
One of the two solutions in $\S {SU(6)}$ ($\S {G_2 \times SU(3)}$) 
comes from $\overline{\mc^W \cap \S {SU(3) \times SU(3)}}$, the other one
belongs to a different  irreducible component containing 
 a single point. The decomposition
of $\mc$ into irreducible components is
\begin{equation} \label{dec-iii}
\mc = \{(p,q=p(p-\a)) \} \cup \{ (\a /2,0) \} \cup \{ (2 \a /3, \a ^2 /9) \}.
\end{equation}

\begin{center}
\setlength{\unitlength}{1mm}
\begin{picture}(110,90)(-20,-50)
\put(-20,0){\vector(1,0){40}}
\put(0,-10){\vector(0,1){40}}
\put(20,-2){$p$}
\put(2,30){$q$}
\put(-20,30){{\it a)}}
\put(40,30){{\it b)}}
\put(-20,-20){{\it c)}}
\put(40,-20){{\it d)}}
\put(-17,0){\line(1,0){34}}
\curve(-10,25,0,0,10,25)
\put(9,20){ ${}^{\Sigma_{G_2 \times SU(3)}}$ }
\put(0,0){\circle*{.8}}
\put(-4,-2){${}_{\Sigma_{SO(13)}}$}
\put(-20,0){${}^{\Sigma_{SU(6)}}$}
\put(40,0){\line(1,0){40}}
\curve(49,30.25,60,0,71,30.25)
\put(69,20){ ${}^{\Sigma_{G_2 \times SU(3)}}$}
\put(60,0){\circle*{.8}}
\put(56,-2){${}_{\Sigma_{SO(13)}}$}
\put(40,0){${}^{\Sigma_{SU(6)}}$}
\put(-20,-50){\line(1,0){40}}
\curve(-11,-19.75,0,-50,11,-19.75)
\put(-11.5,-24){ ${}^{\Sigma_{G_2 \times SU(3)}}$}
\put(0,-50){\circle*{.8}}
\put(-5,-52){${}_{\Sigma_{SO(13)}}$}
\put(-20,-50){${}^{\Sigma_{SU(6)}}$}
\put(40,-50){\line(1,0){40}}
\curve(49,-19.75,60,-50,71,-19.75)
\put(69,-30){ ${}^{\Sigma_{G_2 \times SU(3)}}$}
\put(60,-50){\circle*{.8}}
\put(50,-52){${}_{\Sigma_{SO(13)}}$}
\put(40,-50){${}^{\Sigma_{SU(6)}}$}
\linethickness{.3mm}
\put(67,-10){\line(0,1){40}}
\put(68,-7){${}_{{\mc^W}_{(1)}}$}
\put(60,0){\circle*{1.2}}
\curve(-10,-25,0,-50,10,-25)
\curve(-12.6,-20.5,0,-60.75,12.6,-20.5) 
\put(11.5,-30){${}_{{\mc^W}_{(1)}}$}
\put(-10,-30){ ${}^{{\mc^W}_{(2)}}$}
\curve(56.5,-20.25,62.5,-56.25,68.5,-20.25)
\put(62.5,-50){\circle*{1.2}}
\put(63.3,-47.2){\circle*{1.2}}
\end{picture}
\vspace{1cm}\\
\end{center}

\noindent
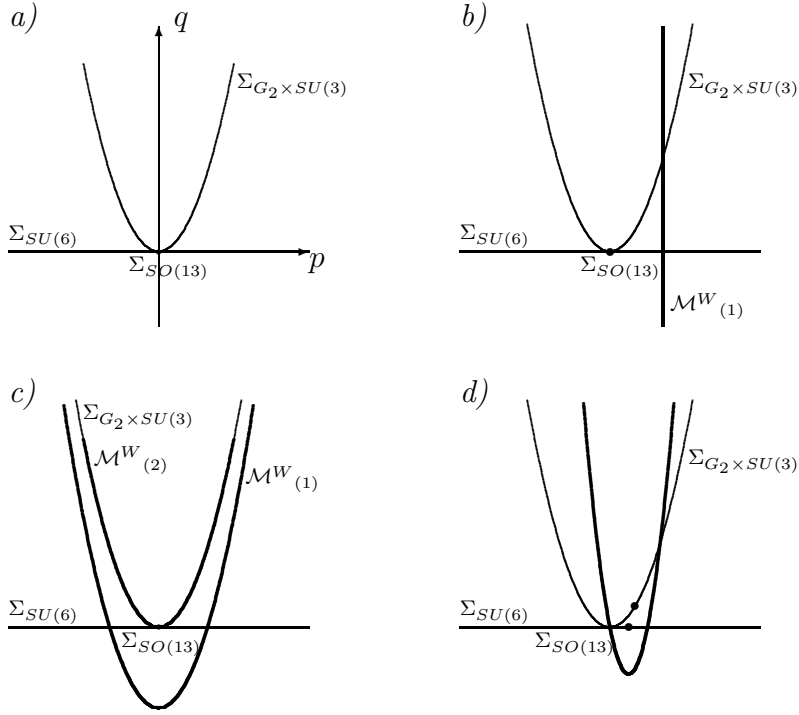
{\it Figure 1: a) The real section $(p,q) \in {\Bbb R}^2$
of the moduli space $\c^2$ of the $SO(13)$ theory with a spinor analyzed in
example~\ref{comp}.1. The figure shows the strata
$\S {G_2 \times SU(3)}, \S {SU(6)}$ and $\S {SO(13)}$, removing
them from the plane we obtain the principal stratum $\S {SU(3) \times SU(3)}$.
b) Moduli space of example~\ref{comp}.1(i),
assuming $f'(p)$ has a single (real positive) root, in which
case $\mc^W$ has two irreducible components, the  line
${\mc^W}_{(1)}$ and the point ${\mc^W}_{(2)}= \S {SO(13)}$.
c) The two
irreducible components of the moduli space of example~\ref{comp}.1(ii)
are parabolas, one of them agrees with the stratum $\S {G_2 \times SU(3)}$.
d) The three irreducible components of the moduli space of
example~\ref{comp}.1(iii) are a parabola and two isolated points,
one of them lying on $\S {G_2 \times SU(3)}$, the other  on
$\S {SU(6)}$ \vspace{.5cm}.}\\

\subsubsection{Integrating out heavy fields}
The procedure described above simplifies if $\mc ^W$ is
known a priori to be irreducible: order the strata as in [i], then 
look for solutions in the first column, then the second one, etc, until 
solutions are found. If this first happens at $\S {(H)}$ and
the solution set is $s \subseteq \S {(H)}$, then
$\mc ^{W} = \overline{s}$. As an application, consider
the problem of identifying composites made heavy by a mass
superpotential $\hw_{mass} = m \hp$, a first step in the process
of integrating out fields from an effective superpotential~\cite{out,susy}.
The set $\c^n_{W_{mass}}$ of critical points of $W_{mass}(\phi) =
\hw_{mass}(\hp(\phi))$ is a vector space, therefore an irreducible
$\c^n$ algebraic subset, and so is $\mc ^{W_{mass}} = \pi(\c^n_{W_{mass}})$.
If $\S {(H)}$ is {\em the} highest dimensional stratum
intersecting $\mc ^{W_{mass}}$, then $\mc ^{W_{mass}} =
\overline{ \{ \hp \in  \S {(H)}| dW^{mass}_{(H)}(\hp)=0 \} }$.\\

\noindent
{\it Example \ref{comp}.2:}
Consider $\hw = m M^F_F$ in $F < N$ SQCD (refer to Example \ref{ls}.1).  
There are no solutions at the main stratum $\S {SU(N-F)} = 
{\Bbb M}^F_F$, the set of rank $F, F \times F$ matrices.
We look for solutions at the only stratum
in the second column, which is 
$\S {SU(N-F+1)} = {\Bbb M}^F_{F-1}$. We use  Lagrange multipliers 
and look for critical points of $m M^F_F + \a \; \text{det} \; M$ 
satisfying cofactor $M \neq 0$.
The solution set is $\mc ^W \cap \S {SU(N-F+1)}
= \{ M | M = \text{diag} \; (M_L,0)  \; M_L \in {\Bbb M}^{F-1}_{F-1} \}
\simeq  {\Bbb M}^{F-1}_{F-1}$, taking its  closure
we obtain $\mc^W = \{  M | M = \text{diag} \; (M_L,0) \} = 
{\Bbb M}^{F-1}$.
This tells us that the heavy  fields are
$M^F_i$ and $M^i_F, i = 1,...,F$. \\

In the special case of an irreducible $\mc ^W$ intersecting the 
main stratum $\S {(G_P)}$ all we need to know are the constraints defining
$\mc = \overline{ \S {(G_P)}}$, as these are the ones
used in the Lagrange multiplier method.
method.\\

\noindent
{\it Example \ref{comp}.3:}
$W=0$,  $N=2, F=3$ SQCD contains six $SU(2)$ fundamentals 
$Q^{\a}_i, i=1,...,6$. The basic invariants are 
$V_{ij} = Q^{\a}_i Q^{\b}_j \epsilon_{\a \b}$. 
The  moduli space is $\mc = 
\{ V | \epsilon^{i_1i_2i_3i_4i_5i_6} V_{i_1i_2} V_{i_3i_4} = 0 \}$ 
and has two strata: $\S {1} = \{ V \in \mc | V \neq 0 \}$, 
and $\S {SU(2)} = \{ V=0 \}$.
The quantum theory 
develops the effective superpotential $\hw _{eff} = 
\epsilon^{i_1i_2i_3i_4i_5i_6} V_{i_1i_2} V_{i_3i_4} V_{i_5i_6}
/ \Lambda_{(F=3)}^3$, $\mc$ is the set of stationary points
of $W_{eff}$.
 Adding a tree 
level superpotential $\hw=m V_{56}$ and integrating out the heavy 
composite fields $V_{5i}, V_{6i}$ from $\hw_{eff}+\hw_{tree}$
we obtain  the quantum deformed $F=N=2$ moduli 
space $Pf \; V = \Lambda_{(F=2)}^4$.
Suppose we want a ``low energy description" of
the integrating out procedure.
We do not know the elementary quark composition
of the $V_{ij}$'s and need to find out
 which
fields are made heavy by $\hw = m V_{56}$.
Following the above recipe, we first look for  the set stationary points
 of the restriction of $W_{tree}$ to the main stratum of $\mc$,
then  take the closure of the solution set.
The stationary points of 
$m V_{56} + \epsilon^{i_1i_2i_3i_4i_5i_6} V_{i_1i_2} V_{i_3i_4} 
\lambda_{i_5 i_6}$ ($\lambda_{ij}= -\lambda_{ji}$ are Lagrange multipliers)
satisfy the following conditions: 
$\lambda \neq 0, \lambda_{5i}=\lambda_{6i}=0; 
V \neq 0, V_{5i}=V_{6i}=0, \epsilon^{i_1i_2i_3i_4 5 6} V_{i_1i_2}
 V_{i_3i_4} = 0$, and $\epsilon^{i_1i_2i_3i_456} V_{i_1i_2}\lambda_{i_3i_4} 
= -m/2$. We conclude the light fields are $V_{ij}, i,j \neq 5,6$,
classically constrained by $\epsilon^{i_1i_2i_3i_4 5 6} V_{i_1i_2}
 V_{i_3i_4} = 0$.
Thus, the fields to integrate out are $V_{5i}$ and $V_{6i}$.

\subsubsection{Potentials lifting  flat directions}
The fact that $\mc^W \cap \S {(H)}$ is the set of
stationary points of ${\hw}_{(H)}$ can
 be applied to  a systematic search of
superpotentials $\hw$ lifting the non trivial classical flat directions of
a theory with given gauge group $G$ and matter content $\phi$.
The interest in finding superpotentials satisfying this
condition lies in the fact that the
resulting theory  is a candidate for
dynamical supersymmetry breaking~\cite{dsb}.
If the theory contains no singlets, $d {\hw}_{(G)}=0$ is trivially
satisfied, since $\S {(G)}$ is  zero dimensional, and the problem
in hand is finding all $\hw$ for which the equation
 $d {\hw}_{(H)}=0$ has no solution if $(H) < (G)$. \\

\noindent
{\it Example \ref{comp}.4:}
Let us look for all
superpotentials lifting flat directions
in the $SO(13)$ with a spinor theory above,  which are
at most quadratic in the invariants $(p,q)$\footnote{Note that
there is no renormalizable gauge invariant superpotential for this theory,
since $p=S^4$ and $q=S^8$, $S$ the spinor field.},
$\hw = Ap+Bq+Cp^2/2+Dq^2/2+Epq$.
We have
\begin{eqnarray}
{\hw}_{SU(6)} &=& Ap+Cp^2/2, p \neq 0, \label{6} \\
{\hw}_{G_2 \times SU(3)} &=& Ap + (B/4+C/2)p^2+Ep^3/4+Dp^4/32, p\neq 0
\label{g2}.
\end{eqnarray}
There are two possibilities: \\
(i) The complex polynomial  $Ap + (B/4+C/2)p^2+Ep^3/4+Dp^4/32$ has
no zeroes, then $B+2C=D=E=0, A \neq 0$. The condition that
$d{\hw}_{SU(6)}/dp =(A+Cp)$ has no $p \neq 0$ zeroes adds  $C=0$, then
$\hw = Ap$ and $d{\hw}_{SU(3) \times SU(3)}$ is  automatically non-zero.\\
(ii)  The polynomial $A + (B/2+C)p + 3Ep^2/4+Dp^3/8$
has zero as its  only root, then $A=0$ and only one of
of $B+2C, E$ or $D$ is non-zero. Adding $d{\hw}_{(H)} \neq 0$ for
$H=SU(6)$ and $SU(3) \times SU(3)$ gives $A=E=D=0$, $B,C$ and $B+2C$
non-zero. \\
In conclusion, the only superpotentials at most quadratic in the invariants
that lift all classical flat directions are $\hw = Ap$ and $\hw = Bq+Cp^2/2$ with
$B,C,$ and $B+2C$ all different from zero. \\

\noindent
{\it Example \ref{comp}.5:}
Consider the $SU(3) \times SU(2)$
model of Affleck, Dine and Seiberg~\cite{dsb}. The matter content is
a field $Q$ in the $({\bf 3}, {{\bf 2}})$,
 fields $\overline{u}$ and $\overline{d}$ in the
$(\overline{{\bf 3}}, {\bf 1})$
and a field $L$ in the $({\bf 1}, {\bf 2})$. The basic invariants
are $x^1=Q\overline{u}L, x^2=Q\overline{d}L$ and $x^3=Q\overline{u}
Q\overline{d}$. They are
unconstrained, then $\mc = \c^3$.
The strata are readily seen to be $\S {1}~=~\{ (x^1,x^2,x^3) | x^3 \neq 0 \}$,
$\S {SU(2)}~=~\{ (x^1,x^2,x^3) | x^3=0 \text{ and } (x^1,x^2) \neq (0,0) \}$,
and $\S {SU(3) \times SU(2)}~=~\{(0,0,0)\}$. Assume $\hw$ is less than cubic
in the composites, $\hw = A_i x^i + B_{ij} x^i x^j/2$. The
supersymmetric vacua in $\S {1}$ and $\S {SU(2)}$   are
respectively the solutions to
the equations
\begin{eqnarray}
d {\hw}_{1} &=& B_{ij}x^j + A_i = 0, x^3 \neq 0,   \label{c1} \\
d {\hw}_{SU(2)} &=& B_{i'j'}x^{j'} + A_{i'} = 0, (x^1,x^2) \neq (0,0), \label{c2}
\end{eqnarray}
 where $i,j=1,2,3$ and $i',j'=1,2$.
Requiring that $\hw$ lifts all non trivial flat points is equivalent to
demanding that the only possible solution to the linear system
in~(\ref{c1})
be the trivial one~\footnote{Any $x^3=0$ solution would also be a solution
of equation (\ref{c2}) unless $x^1=x^2=0$.} and also that the
only possible solution of the linear system in~(\ref{c2}) be trivial.
This leads to the following three possibilities:
(i) neither $B_{ij}x^j + A_i = 0$ nor  $B_{i'j'}x^{j'} + A_{i'} = 0$
has a solution,
(ii) $B_{ij}x^j + A_i = 0$ has no solution and
$B_{i'j'}x^{j'} + A_{i'} = 0$ only for $(x^1,x^2)=(0,0)$,
which implies $A_1=A_2=0$ and det $(B_{i'j'}) \neq 0$; and
(iii) each linear system has the trivial solution as the only one,
i.e, $A_i=0$, det $(B_{ij})
\neq 0$ and det $(B_{i'j'}) \neq 0$. As an example,
$B_{ij}=0$ and  $(A_1,A_2) \neq (0,0)$ is a possible solution,
and choosing $A_3=0$ we obtain the only renormalizable gauge invariant
superpotential lifting all flat directions.\footnote{The
 Affleck, Dine and Seiberg theory corresponds to the choice $B_{ij}=0$,
$A_2=A_3=0$.} A
$B_{ij} \neq 0$ example is $\hw = Bx^1x^2+Cx^3$.\\

\subsubsection{Patterns of gauge symmetry braking in $W \neq 0$ theories}

Theorem~I.a,b,c gives a well defined pattern for the breaking
of the  gauge symmetry $G$ in theories with zero superpotential. There is an
order relation in the set ${\Bbb S}$ of  (classes of) unbroken
 subgroups of $G$
at different vacua,
namely $(H) < (H')$ if $H$ is conjugate to
a proper subgroup of $H'$. ${\Bbb S}$
contains a unique
maximal class  $(G)$ and a unique minimal isotropy
group $(G_P)$, and, when ${\Bbb S}$ is  arranged as
explained at
the beginning of Section~(\ref{exs}),  all patterns of
gauge symmetry breaking of the $W=0$ theory from
 $G$ to  $G_P$
are exhibited.
If a superpotential $W$ is
turned on, the resulting moduli space will intersect {\em some} of the
strata $\S {(H)}$ of the $W=0$ theory. From the stratification
$\mc = \cup_{(H)} \S {(H)}$ of $\mc$, and the fact that
$\mc^W \subset \mc$, we obtain
the stratification of $\mc^W$:
\begin{equation} \label{ws}
\mc^W = \bigcup_{(H) \in \ws} \left( \mc^W \cap \S {(H)} \right),
\end{equation}
$\ws$ being the set
 of (classes of) unbroken subgroups at vacua in the theory with
superpotential $W$, i.e., the set of strata intersecting $\mc^W$.
As $W$ lifts flat directions,
some of the unbroken subgroups of the $W=0$ theory are
missing in $\ws$. The partial order relation in ${\Bbb S}$
is inherited by $\ws$,
this is used to order the $\mc^W$  strata $\mc^W \cap \S {(H)}$.
It is then natural to ask
if some of the conditions in Theorem~Ia,b,c subsist in the theory
with superpotential. Consider first Theorem~I.a,  the
stratification~(\ref{ws}) is finite, but it is easy to see that, generically,
the strata are not manifolds. Consider e.g.  the $SO(13)$ theory
with a spinor of Example~\ref{comp}.1 with   a superpotential
$\hw(p,q) = (p-p_0)^2(q-q_0)^2, q_0 \neq 0, p_0^2/4$. The $SU(3) \times SU(3)$
stratum of this theory, being singular at $(p_0,q_0)$, is not a manifold.
 Point (b) in Theorem~I
does not hold if $W \neq 0$, the three superpotentials in Example.~\ref{comp}.1
illustrate this fact. Most important, point (c) in Theorem~I
is no longer true  either. Generically, the set
of minimal  (classes of) unbroken  subgroups contains more
than one element.
A simple example is  the $SO(13)$
 theory with a spinor
 and superpotential $\hw(p,q)=q(q-p^2/4)$, which
exhibits
the following pattern of symmetry breaking:
\begin{equation} \label{p1} \begin{array}{ccc}
& & G_2 \times SU(3)  \\
 & \diagup &    \\
SO(13) &  &    \\
   & \diagdown  &\\
&& SU(6)  \end{array} \end{equation}
Although dim $G_2 \times SU(3) < \text{ dim } SU(6)$, $G_2 \times SU(3)$
is not conjugate to an $SU(6)$ subgroup, there is no Higgs flows between
these two unrelated theories.
A unique maximal unbroken gauge subgroup (minimal stratum)
exists if the theory  contains no $G$ singlets, this is $(G)$ ($\S {(G)}$).
 Yet, theories with a gauge singlet may not even
have a maximal unbroken gauge subgroup when a superpotential is
turned on. As an example, add  an $SO(13)$  singlet
 $r$ to the $SO(13)$
theory with a spinor. The moduli space is $\mc = \{(p,q,r)\} = \c^3$ and the
strata are the sets of $(p,q,r)$ constrained by the same  equations
in (\ref{ssspin13}). Take $\hw(p,q,r)=r(p-p_0), p_0 \neq 0$,
then $\mc^W$ is the line $\{ (p_0,q,0), q \in \c \}$   which does not intersect
$\S {SO(13)} = \{ (0,0,r) \}$. The pattern of gauge symmetry breaking
of this theory,
\begin{equation} \label{p2} \begin{array}{ccc}
G_2 \times SU(3) & &  \\
 & \diagdown &    \\
 &  &  SU(3) \times SU(3)  \\
   & \diagup  &\\
SU(6) &&   \end{array} \end{equation}
has {\em two} maximal $SO(13)$ subgroups (minimal strata)
from where to start flowing  down to smaller subgroups by Higgs mechanism.
The reader can check that the superpotential $\hw = q(q-p_0^2/2)+r(p-p_0)^2,
p_0 \neq 0$ lifts all $SO(13)$ and $SU(3)\times SU(3)$ vacua, then the moduli
space of this theory has two maximal (minimal) unbroken gauge subgroups.\\
The situation gets better if we consider instead
{\em irreducible components} ${\mc^W}_{(i)} \subseteq \mc^W$.
According to the results in Appendix~A, there is a unique maximal
stratum $\S {(H_i)}$
intersecting ${\mc^W}_{(i)}$ and
equation~(\ref{icms}) holds. This is analogous to 
equation~(\ref{closure}) in Theorem~I.b when
applied to the maximal stratum (only).
Irreducible moduli spaces share this important
property with  the $W=0$ (irreducible) moduli spaces.\\
The results in Section~\ref{comp} are gathered below. \\

\noindent
{\bf Corollary 1 of Theorem~I:}  Let $\hp ^i(\phi),
i=1,...,s$, be
a basic set of holomorphic
$G$ invariants of the theory  with  matter
content $\{ \phi \}$ and gauge group $G$,
$p_{\a}(\hp(\phi))\equiv 0$ the algebraic constraints among the basic
invariants, $\mc = \{ \hp \in \c^s | p_{\a}(\hp)=0 \}$ the
moduli space of the $W=0$ theory.
 Let $\S {(H)} \subseteq \mc$  be the stratum of
vacua with (classes of) unbroken  gauge subgroups conjugate to $H \subseteq G$,
$K_{\b}^{(H)}(\hp)=0$ the polynomial equations defining (the closure of)
$\S {(H)}$.  Let $W(\phi) =
\hw (\hp (\phi))$, be a superpotential and
$\hw _{(H)}$ the restriction of $\hw$ to the complex manifold
$\S {(H)}$.
\begin{itemize}
\item [(a)] The set of vacua in $\S {(H)}$, $\mc ^W \cap \S {(H)}$,
is the set of
critical points $d \hw _{(H)}~=~0$~\cite{plb,as}.
This can be obtained (i) by covering the complex manifold
$\S {(H)}$ with local coordinates $x^i$ and solving
$\partial \hw_{(H)}(x) / \partial x^i =0$, or (ii) by using Lagrange
multipliers to find the stationary points of
$\hw(\hp)+ C^{\b} K^{(H)}_{\b}(\hp)$, and then discarding the
solutions not in $\S {(H)}$.
\item [(b)] Generically, if $W \neq 0$
the strata $\S {(H)} \cap \mc^W$ are {\em not}
manifolds, $\overline{\mc^W \cap \S {(H)}} \neq  \cup_{(L) \geq (H)}
(\mc^W \cap \S {(L)})$, and the sets of maximal
and minimal classes of unbroken gauge subgroups contain more than one
element.
\item [(c)] If $\mc^W = \cup_i {\mc^W}_{(i)}$ is the decomposition of
$\mc^W$ into irreducible components, then for each $i$ there is a maximal stratum
$\S {(H_i)}$ intersecting ${\mc^W}_{(i)}$, and
${\mc^W}_{(i)} = \overline{{\mc^W}_{(i)} \cap \S {(H_i)}}$.
\end{itemize}

\subsection{Massless fields after Higgs mechanism}
\label{mf}

The differential $\pi'_{\phi_0}$ of the
 map $\pi: \phi \to \hp(\phi)$ at the \df point $\phi_0$
is given by the matrix
$\partial \hp^i(\phi_0) / \partial \phi^j, \;
\pi'_{\phi_0}: \delta \phi^j \to  \delta \hp^i =
(\partial \hp^i(\phi_0)/ \partial \hp^j)
\delta \hp^j$.
Note that $\pi: \c^n \to \mc = \{ \hp \in \c^s | p_{\a}(\hp)=0 \}$,
then
$\pi'_{\phi_0}: T_{\phi_0}\c^n \to T_{\hp_0} \mc$, $\hp_0 \equiv
\hp(\phi_0)$.
 The tangent at $\phi_0$ of $\c^n$ is $T_{\phi_0}\c^n \simeq \c^n$,
and the tangent $T_{\hp_0}\mc$
is the space of moduli $\delta \hp$ consistent with the linearized
constraints,  $(\partial p_{\a}(\hp_0)/
\partial \hp^j) \delta \hp^j = 0$ (assuming the constraints satisfy
the requirement in footnote~2.)
A natural question to ask is whether $\pi'_{\phi_0}$ 
makes  $T_{\hp_0}\mc^W
\subseteq T_{\hp_0} \mc$
 isomorphic to the space of massless modes at a supersymmetric vacuum
  $\phi_0$ in the classical regime.
We devote this section to answering this question.\\
\underline{$W=0$ case:} \\
 The  space $\{ \delta \phi \} = T_{\phi_0} \c^n = 
  \t {\phi} \oplus \n {\phi} \oplus \s {\phi}$, $\delta \phi$ uniquely 
decomposes as $\delta \phi = \delta t + \delta n + \delta s$.
 The fields $\delta t$ in
 $\t {\phi}$ are eaten by the broken gauge
generators (two real fields per heavy vector superfield).
Thus, if  $W=0$,  
the light fields in unitary gauge, i.e., the massless fields
after Higgs mechanism (MFHM) are  those in $\n {\phi} \oplus \s {\phi}
\equiv {\rm NMFHM} \oplus {\rm SMFHM}$, where (N)SMFHM is a short for
(non)singlet massless fields after Higgs mechanism.
According to Theorem~I.e $\pi'_{\phi_0}$ annihilates $\n {\phi_0}$,
 {\em the {\rm NMFHM} are not represented
in $T_{\hp_0} \mc$}. On the other hand, the rank of $\pi'_{\phi_0}$
is not the whole $T_{\hp_0}\mc$ but the tangent to the stratum $\S {(G_{\phi_0})}
\equiv \S {\hp_0}$ through $\hp_0$, and so {\em there are spurious fields
$C_{\hp_0}  \subseteq T_{\hp_0}\mc$, unrelated to the} MFHM.
The situation is illustrated in the following diagram:
\begin{eqnarray}  \label{w=0}  \begin{array}{ccccc}
T_{\hp_0}\mc  &=& T_{\hp_0} \S {\hp_0}& \oplus &  C_{\hp_0} \\
    &&  \, \| & &\\
 {\rm MFHM} &=& \s {\phi_0}& \oplus &\n {\phi_0}\\ \end{array}
\end{eqnarray}
We would like to know when  $C _{\hp_0}$ and
$\n {\phi_0}$ are null.
 We consider separately the following
two cases: \\
(i) $\hp_0 \in \S {(G_P)}$ ($\S {\hp_0} = \S {(G_P)})$:
From Theorem~I.b-c  $\mc = \overline{\S {\hp_0}}$, then $T_{\hp_0}
\S {\hp_0} = T_{\hp_0}\mc$ and  $C_{\phi_0}$ is null.
From Theorem~I.f $\n {\phi_0}$ is null if and only if the theory
is stable. \\
(ii) $\hp_0 \notin \S {(G_P)}, (\S {\hp_0} < \S {(G_P)})$:
From Theorem~I.b  $\S {\hp_0}$ lies in the boundary
of the principal stratum, dim~$\S {\hp_0} < \text{dim }
\S {(G_P)} = \text{dim } \mc \leq \text{dim } T_{\hp_0} \mc$, and so
$T_{\hp_0} \S {\hp_0} \subsetneq T_{\hp_0} \mc$,  $C _{\hp_0}$ is
non trivial. In this case  also  $(G_{\phi_0}) > (G_P)$
i.e., $G_P$ is conjugate to a proper subgroup of $G_{\phi_0}$,
as follows from the definition of the order relation among
strata and isotropy classes, and so dim~$G_{\phi_0} > \text{ dim }
G_P$. We can use this information together with
Theorem~I.e to show that $\n {\phi_0}$
is not null. Pick any \df point  $\phi_1$ such that
$\hp(\phi_1) \in \S {(G_P)}$, then (see footnotes~1 and 2)
\begin{eqnarray} \label{dn0}
\text{dim } \n {\phi_0} &=& n - (\text{dim}_{\Bbb R} G -
\text{dim}_{\Bbb R}  G_{\phi_0}) -  \text{dim } \S {(G_{\phi_0})}\\
\nonumber & > &  n - (\text{dim}_{\Bbb R} G - \text{dim }_{\Bbb R}
G_P) - \text{dim } \S {(G_P)} = \text{dim } \n {\phi_1} \geq 0,
\end{eqnarray}
 In other words,
Higgs mechanism at a
vacuum $\phi_0$ with $(G_{\phi_0}) > (G_P)$
always leaves a theory with fields transforming non trivially under
$G_{\phi_0}$.\\
In conclusion, for any $W=0$ theory, spurious fields in $T_{\hp_0} \mc$
are always present unless $\hp_0$ belongs to
the principal stratum.
$\pi'_{\phi_0}$ is an isomorphism between the space of SMFHM
and $T_{\hp_0} \S {\hp_0}$. The NMFHM are unseen as
moduli $\delta \hp$, they are
always present, except at  vacua in the principal stratum of a stable theory. \\
\underline{Generic $W$ case:} \\
The
space of massless fields
 at the supersymmetric vacuum $\phi_0$
is the kernel of $W_{ij}(\phi_0) = \partial^2 W(\phi_0)/\partial \phi^i
\partial \phi^j$. The kernel includes the
eaten fields $\t {\phi_0}$, as follows from the $G^c$ invariance of $W$
\begin{equation}
 W_i(\phi)T^i_k \phi^k = \frac{d}{ds}{\Bigg\vert_{s=0}} W(e^{sT}\phi) = 0,
 \;\;\;
\forall \; T \in \lie {G^c}, \phi \in \c^n,
\end{equation}
by taking a $\phi$ derivative an using the $F-$flatness of $\phi_0$:
\begin{equation}
0 = \frac{\partial}{\partial \phi^j}{\Bigg\vert_{\phi=\phi_0}} 
W_i(\phi) T^i_k \phi^k = 
W_{ij}(\phi_0) T^i_k \phi_0^k, \;\;\; \forall \; T \in \lie {G^c}.
\end{equation}
As $W_{ij}(\phi_0)$ is $G_{\phi_0}$ invariant, it
cannot mix $\n {\phi_0}$ and $\s {\phi_0}$, otherwise, we could 
write a $G_{\phi_0}$ invariant mass term $W_{ij}(\phi_0) \delta \phi^i \delta 
\phi ^j$ mixing singlets $\delta s$  with non singlets $\delta n$. 
We conclude that, under $\c^n = \t {\phi_0} 
\oplus \n {\phi_0} \oplus \s {\phi_0}$,
 $W_{ij}$ is block diagonal:
\begin{eqnarray}   \label{wij}
&& \begin{array}{ccc} \t {\phi_0} \n {\phi_0} & \s {\phi_0} \end{array} \\ \nonumber
W_{ij}(\phi_0) = \begin{array}{c} {\t {\phi_0}} ^* \\ {{\n {\phi_0}}}{}^* \\ 
{\s {\phi_0}} ^* \end{array} && 
 \left( \begin{array}{ccc} 0 &0&0 \\ 0 & N_{ij} & 0 \\ 0&0&
S_{ij} \end{array} \right)  
\end{eqnarray}
After Higgs mechanism we are left with $\n {\phi_0} \oplus \s {\phi_0}$
and so MFHM~$= \text{ker } S_{ij} \oplus \text{ ker } N_{ij}
\equiv {\rm SMFHM} \oplus {\rm NMFHM}$. We consider the SMFHM space first.
In view of equation~(\ref{w=0}), $\pi'_{\phi_0}$ makes
$\s {\phi_0}$ isomorphic to $T_{\hp_0} \S {\hp_0}$.
From this isomorphism and  the inverse function theorem
follows that  a  neighborhood of the origin of $\s {\phi_0}$
can be used as a coordinate patch of the complex
manifold $\S {\hp_0}$ around
$\hp_0$. Note that if $x^j$ and $y^k$ are any two
local coordinate sets of $\S {\hp_0}$
with $x=y=0$ at $\hp_0$, and $\hp_0 \in \mc^W$, then $\partial
{\hw}_{(G_{\phi_0})} / \partial y^k = 0$ at $y=0$ (Corollary 1.a in
Section~(\ref{comp})), and
\begin{equation}
[\hw_{(G_{\phi_0})}]_{ij}(\hp_0) \equiv
\frac{\partial^2 \hw _{(G_{\phi_0})}}{\partial x^i \partial x^j}
{\Bigg\vert_{x=0}}  =
\left( \frac{\partial^2 \hw _{(G_{\phi_0})}}{\partial y^k \partial y^l}
{\Bigg\vert_{y=0}}\right)
\left( \frac{\partial  y^k}{
\partial x^i}{\Bigg\vert_{x=0}} \right) \left(\frac{\partial y^l}{
\partial x^j }{\Bigg\vert_{x=0}}\right)
\end{equation}
transforms as a $(0,2)$ tensor at $\hp_0$
\footnote{This tensor can be written more covariantly as
 $\nabla _i \nabla_j \hw _{(G_{\phi_0})}=
\partial_i \partial_j \hw _{(G_{\phi_0})} + \Gamma^k_{ij} \partial_k
\hw _{(G_{\phi_0})}$,
$\nabla_i$ an arbitrary  covariant derivative on
the manifold $\S {\hp_0}$, as the second term vanishes
when evaluated at a vacuum.}, then
\begin{equation}
\text{ker } [\hw_{(G_{\phi_0})}]_{ij}(\hp_0) = \left\{ \delta x^i \,
| \,\frac{\partial^2 {\hw}_{(G_{\phi_0})}(\hp_0)}{\partial x^i \partial x^j}
\delta x^j =0 \right\}
\end{equation}
is a well defined (coordinate independent)
subspace of $T_{\hp_0} \S {\hp_0}$ with complement $D_{\hp_0}^W$.
This subspace is obtained by linearizing at $\hp_0$ the constraints
$\partial \hw _{(G_{\phi_0)}} / \partial x^i = 0$ defining
$\mc^W \cap \S {(G_{\phi_0})}$ (Corollary~1.a), then is the tangent space
$T_{\hp_0} (\mc^W \cap \S {(G_{\phi_0})})$~\footnote{It might
actually be bigger than  $T_{\hp_0} (\mc^W \cap \S {(G_{\phi_0})})$ if
there is problem of the type indicated in footnote~2. This
may happen if $\hw(\hp)$ is of high degree in the invariants (therefore
non renormalizable), or the constraints defining the strata are
high degree polynomials.}.
 In  the coordinates $\delta s$ of $\S {\hp_0}$,
$[\hw_{(G_{\phi_0})}]_{ij}=S_{ij}$,  the $W \neq 0$
analogous of eq.(\ref{w=0}) is
\begin{eqnarray}  \label{wneq0}  \begin{array}{ccccccc}
T_{\hp_0}\mc  &=& \text{ ker } [\hw_{(G_{\phi_0})}]_{ij} &\oplus &  D_{\hp_0}^W
& \oplus &  C_{\hp_0} \\
    && \| & &&&\\
 {\rm MFHM} &=& \text{ker } S_{ij} & \oplus &\text{ker }  N_{ij} && \\ \end{array}
\end{eqnarray}
Among the MFHM, the SMFHM  ker~$S_{ij} \simeq \text{ ker }
[\hw_{(G_{\phi_0)}}]_{ij}$ are represented as moduli, whereas
the NMFHM ker $N_{ij}$ are not. The moduli in $  D_{\hp_0}^W
 \oplus   C_{\hp_0}$ are spurious.
We establish conditions under which the space ker~$N_{ij}$ of NMFHM
is null:\\
(i) $\hp_0 \in \S {(G_P)}$: If the theory is stable, $\n {\hp_0}$
is null (Theorem~I.f) and so is ker~$N_{ij}$. If the theory is unstable,
$\n {\phi_0}$ is non trivial and the theory with gauge group $G_P =
G_{\phi_0}$ and matter content $\{ \delta n \} = \n {\phi_0}$ has no holomorphic
$G_{\phi_0}$ invariants. In particular, $N_{ij} \delta n^i \delta n^j$,
being holomorphic and $G_{\phi_0}$ invariant,
must be zero, then $N_{ij}=0$ and ker~$N_{ij}=\n {\phi_0}$ is
not null.\\
(ii) $\hp_0 \notin \S {(G_P)}$: According to eq.(\ref{dn0})
 dim~$\n {\phi_0} > 0$. However, no general statement can be made
about ker~$N_{ij} \subseteq \n {\phi_0}$ if $W$ is unknown.
An exception is when  the theory with gauge group $G_{\phi_0}$
and matter content
$\n {\phi_0}$ is known to be chiral (no quadratic holomorphic invariants),
case in which
we can repeat the argument above to show that $N_{ij}=0$
and so ker~$N_{ij}=\n {\phi_0}$ is not null.\\

These results are gathered in the Corollary
below: \\

\noindent
 {\bf Corollary 2  of Theorem~I:}
 The space MFHM of massless fields after Higgs mechanism
at a vacuum with residual gauge group $H$ is the direct sum of
the $H$ singlet space SMFHM and the non-singlet space NMFHM
\begin{itemize}
\item [(a)] Let $x^i$ be any set of local coordinates
of $\S {(H)}$ around a vacuum $\hp_0$. SMFHM  is isomorphic
to the subspace $\{ \delta x^i \,  |
(\partial^2 \hw_{(H)} (\hp_0)/
\partial x^i \partial x^j)\ \, \delta x^j = 0 \} \subseteq T_{\hp_0} \S {(H)}$.
 ${\rm SMFHM} = T_{\hp_0}(\mc ^W \cap \S {H})$ (see however footnote~8).
\item [(b)] The NMFHM
are annihilated by $\pi'_{\phi_0}$, and so they are
missing (unseen as moduli $\delta \hp$) in the moduli space.
 For any $W$, this set
is trivial if $\hp_0$ belongs to the principal stratum of a stable
theory, non-trivial if $\hp_0$ is in the principal stratum of an unstable theory.
\item [(c)] At vacua in non principal strata there are (potentially) missing
NMFHM if $W=0$ ($W \neq 0$).
\end{itemize}

\noindent
{\it Example \ref{mf}.1.} Coming back to Example~\ref{lec}.2, at
$\S {SO(N-1)} ^{(A)}$ is  $\hw_{SO(N-1)} = my^2/x, x \neq 0$,
then the vacuum
condition $d \hw _{SO(N-1)} = (-my^2/x^2,2my/x) = 0$ implies $y=0$
and
\begin{equation}
\left( \hw_{SO(N-1)}\right)_{ij} = \frac{2m}{x} \left( \begin{array}{cc}
y^2/x^2 & -y/x \\ -y/x & 1 \end{array} \right) =
\frac{2m}{x} \left( \begin{array}{cc}
0 & 0 \\ 0 & 1 \end{array} \right),
\end{equation} 
giving  a single massless $SO(N-1)$ singlet after Higgs mechanism,
a fact that can
be readily verified in a microscopic field description.\\

\noindent
{\it Example \ref{mf}.2}
  We continue the analysis of
the three different cases of Example~\ref{comp}.1. \\
{\em (i) $ \hw (p,q) = f(p) \equiv (p-p_0)^2, p_0 \in  {\Bbb R}^{>0}$
(Figure 1.b).} \\
Using coordinate charts as in Example~\ref{comp}.1
we  get
\begin{eqnarray} \label{rank}
[ \hw_{SU(3) \times SU(3)} ]_{ij} &=&\text{diag }~(f''(p), 0), \\ \nonumber
[\hw_{SU(6)}] _{ij} &=& f''(p),   \\ \nonumber
[\hw_{G_2 \times SU(3)}] _{ij} &=& f''(p),
\end{eqnarray}
the dimensions  of the SMFHM space at $SU(3) \times SU(3),
G_2 \times SU(3), SU(6)$ and $SO(13)$ vacua equal
$1,0,0$ and $0$ respectively. Note that there is no problem of the kind
mentioned in footnote~8. We can use Corollary~2.a, SMFHM $ =
T_{\hp_0} ( \mc^W \cap \S {(G_{\phi_0})})$, and  the dimension of SMFHM
can easily be read off from fig.~1.b.
At the $\hp=0$ vacuum we have
the original theory, for which the space of SMFHM is null,
that is why dim~$T_{0}(\S {(G)} \cap \mc^W)=$
dim~$\S {(G)}=0$. The (real section) $(p,q) \in {\Bbb R}^2$
of the component $p=p_0$ of $\mc^W$
is a vertical line intersecting all strata but $\S {SO(13)}$ (fig.1.b).
The line intersects $\S {SU(3) \times SU(3)}, \S {G_2 \times SU(3)}$, and
$\S {SU(6)}$ at sets of dimension $1,0$ and $0$, these are
the dimensions of the SMFHM spaces for vacua in
these strata. All vacua in the main
stratum have a null NNMFHM space,   because  the theory
is stable. At vacua in  smaller strata there
could be NMFHM,  unseen
as moduli $\delta \hp$. \\
{\em (ii) $ \hw (p,q) = (p^2-4q-m^8)^2/M^{13}$ (Figure 1.c).}\\
 We use  again Corollary~2
to  read from  figure~1.c
the dimension of the SMFHM space at each vacuum.
$\mc^W$  has two irreducible components:
$\mc^W = {\mc^W}_{(1)} \cup {\mc^W}_{(2)}$, the two parabolas
in figure~1.c. Although ${\mc^W}_{(1)}$
is  one dimensional,
its intersection with $\S {SU(6)}$ is zero dimensional,
and so there is a single  massless singlet at each $SU(3) \times SU(3)$
vacuum in ${\mc^W}_{(1)}$,  no massless $SU(6)$ singlet
at any of the two $\S {SU(6)}$ vacua. A similar analysis holds for
the one dimensional manifold
${\mc^W}_{(2)}$. ${\mc^W}_{(2)} \cap \S {G_2 \times SU(3)} =
{\mc^W}_{(2)} \setminus \{ \hp=0 \}$ is one dimensional,
whereas ${\mc^W}_{(2)} \cap \S {SO(13)} = \{ \hp =0 \}$ is zero dimensional.
Correspondingly, SMFHM is one (zero)
dimensional for ${\mc^W}_{(2)}$ vacua with residual $G_2 \times SU(3)$
($SO(13)$) gauge symmetry.\\
{\em (iii) $ \hw (p,q) = [p(p-\a)-q]^2/M^{13}$ (Figure 1.d).}\\
 Refer to  figure~1.d. The moduli space
has three irreducible components: a parabola ${\mc^W}_{(1)}$
intersecting all four strata, a one point component ${\mc^W}_{(2)}$
in $\S {SU(6)}$ and a single vacuum component ${\mc^W}_{(3)}$
with residual gauge symmetry $G_2 \times SU(3)$.
Every vacuum in ${\mc^W}_{(1)}$ has a one dimensional
space of massless singlets except for the three
vacua with residual gauge symmetry $G_2 \times SU(3)$,
$SU(6)$ and $SO(13)$, which have
no massless singlets
in their spectra. This is so because ${\mc^W}_{(1)} \cap \S{H}$
is zero dimensional for $H = SO(13), SU(6)$ and $G_2 \times SU(3)$,
whereas ${\mc^W}_{(1)} \cap \S{SU(3) \times SU(3)} = {\mc^W}_{(1)}
\setminus \{ \text{ three isolated points } \}$ is one dimensional.
There are no SMFHM at vacua in the other two components.\\
We should stress here that the results in this section all refer
to the classical regime. Although for theories with a simple gauge group $G$,
matter fields $\phi$ in a $G$ representation with Dynkin index
$\mu$ greater
than the index $\mu_G$
of the adjoint, and   $W=0$ the classical  moduli space $\mc$ and the quantum
moduli space
are equal, it is generally {\em not} true that classical and quantum
spectra of massless fields agree at every
 vacuum $\hp \in \mc$. As
an example, consider the s-confining theories in~\cite{sc}.
These theories have an effective superpotential $W_{eff}(\hp)$
whose set of stationary points is $\mc$.
In the classical theory,  at the  $\hp=0$ vacuum
we have gauge group $G$ and  matter content $\phi$,
without singlets. Quantum mechanically, evidence indicates that $G$ is
completely broken and the massless spectrum are the unconstrained
moduli $\delta \hp$~\cite{sc,susy}. A second $\mu > \mu_G$ example are
the   theories
with a low energy dual~\cite{susy,sei}, they have equal
classical and quantum moduli spaces, but the classical and quantum massless
spectra
are   completely
different.

\section{Conclusions}     \label{conc}
A low energy description of the moduli space
$\mc^W$ of a $W \neq 0$, ${\cal N} = 1$ gauge theory, one
in which $\mc^W$ is constructed entirely
in the space spanned by the basic holomorphic invariants
$\hp$ without knowing their elementary field
content $\hp(\phi)$, is possible.
The construction requires knowledge of  the constraints
among the basic invariants $\hp$ that define
the $W=0$ moduli space $\mc$, and also of the
stratification $\mc = \cup_H \S {(H)}$
 according to the
unbroken gauge subgroups class $(H)$
at different vacua. Some shortcuts are possible when
searching for isolated irreducible components of $\mc^W$,
a fact that  is useful to identify   heavy composite fields
to integrate out from an effective superpotential, and to construct
superpotentials that lift all flat directions,
leaving a candidate theory for   dynamical symmetry breaking.
The stratification of $\mc$, together with the low
energy construction of $\mc^W$, allows a systematic
study of the  patterns of gauge symmetry breaking.
When $W$ is trivial, there is theory with a minimal unbroken
gauge subgroup $G_P$
to which  flow by Higgs mechanism leads
in many different ways. A non zero superpotential, on the contrary,
may leave a set of vacua with no unique minimal unbroken subgroup,
then different Higgs flows end up at different theories.\\
Among the massless fields
after Higgs mechanism (MFHM)
at a vacuum $\hp \in \mc^W$, the singlets (SMFHM)  are
represented by  moduli $\delta \hp$, whereas the non singlet (NMFHM)
are not.
Being gauge invariant, $W(\phi) = \hw (\hp)$.
$\mc^W \cap \S {(H)}$ is the
set of critical points of the restriction $\hw_{\mid_{\S {(H)}}}$
of $\hw$ to the stratum $\S {(H)}$, whereas the space of SMFHM
at a vacuum $\hp \in \S {(H)}$
is the kernel of the tensor $\nabla_i \nabla_j  \hw_{\mid_{\S {(H)}}}$
at $\hp$, $\nabla$
any covariant derivative on the complex manifold $\S {(H)}$.
In looking for critical points $d {\hw}_{(H)} = 0 $
local coordinates on the complex manifold
$\S {H}$ can be used. An alternative is
using Lagrange multipliers, adding to $\hw$
terms containing the polynomial constraints in the definition of
$\S {(H)}$. The Lagrange multipliers method is safe
in all cases.
The space of NMFHM is null
for vacua  in the principal stratum (where the gauge group is broken
to the minimal subgroup $G_P$) of a stable theory. In unstable
theories, on the contrary, even for vacua $\hp$
 at the principal stratum  there
are NMFHM, unseen as moduli $\delta \hp$.
Unstable theories are  characterized by
the impossibility of breaking the complexified
gauge group to a minimum dimension subgroup by a \df configuration.
Another distinguishing feature of unstable theories is that the dimension
of their  $W = 0$ moduli space $\mc$ violates
the rule $\text{dim}\;  \mc = \text{ dim microscopic matter field space }
- \text{ dim gauge group } + \text{ dim } G_P$.
Theories with  matter fields  in a real representation of the
gauge group are  stable, and this is the case for most (but not all)
of the allowed
representations, since they must be anomaly free. Unstable theories, therefore,
are  rare.

\acknowledgments
This work was partially supported by Fundaci\'on Antorchas, Conicet and
Secyt-UNC.
I thank Witold Skiba for useful comments on the manuscript
and for suggesting the  application of the techniques
here introduced to dynamical supersymmetry breaking.

\appendix
\section{Derivation of equation (40)} \label{appen}
Let
\begin{equation} \label{ic}
\mc^W  =  \bigcup_{i} {\mc^W}_{(i)}
\end{equation}
be the decomposition of $\mc^W$ into irreducible
components.
As $\mc$ is the disjoint union of its strata $\S {(H)}$ we have
\begin{equation}  \label{1}
{\mc^W}_{(i)} = \bigcup_{\S {(H)} \in \sigma_i}
 \left( {\mc^W}_{(i)} \cap \S {(H)} \right),
\end{equation}
where $\sigma_i$ is the set of strata intersecting ${\mc^W}_{(i)}$.
Let  $\sigma_i^{max}$ be the subset of maximal
strata in $\sigma_i$, i.e., $\S {(H)} \in  \sigma_i^{max}$
if and only if  any other  stratum
$\S {(H')} \in \sigma_i$ is either smaller than or  unrelated to $\S {(H)}$.
From Theorem~I.b , any stratum in $\sigma_i$ lies in the closure
of a $\sigma_i^{max}$ stratum, then the union of the strata
in $\sigma_i$ equals the union of the closures of the strata in
$\sigma_i^{max}$ and
\begin{equation}
{\mc^W}_{(i)} = \bigcup_{\S {(H)} \in \sigma_i^{max}}
\left( {\mc^W}_{(i)} \cap \overline{\S {(H)}} \right).
\end{equation}
${\mc^W}_{(i)}$ being irreducible means that one of the closed  sets
in the union above contains the others, i.e., there is a
$\S {(H_i)} \in \sigma_i^{max}$ such that
\begin{equation} \label{3}
{\mc^W}_{(i)} = {\mc^W}_{(i)} \cap \overline{\S {(H_i)}}.
\end{equation}
Equation (\ref{3}) implies that $\sigma_i^{max}$ contains
a single element, namely,  $\S {(H_i)}$. In fact, assuming there is a
$\sigma_i^{max} \ni \S {(H)} \neq \S {(H_i)}$ leads to
a contradiction:
\begin{equation} \label{cont}
 \emptyset \neq {\mc^W}_{(i)} \cap \S {(H)} =
{\mc^W}_{(i)} \cap \overline{\S {(H_i)}}
\cap \S {(H)} \; \Rightarrow \;
\overline{\S {(H_i)}}
\cap \S {(H)} \neq \emptyset.
\end{equation}
From equations~(\ref{int}) and (\ref{cont}) we get $\S {(H_i)} > \S {(H)}$,
contradicting the assumption that $\S {(H)}$ is maximal.
We conclude that there is a single maximal element $\S {(H_i)}$
in the set  $\sigma_i$ of strata intersecting the irreducible
component ${\mc^W}_{(i)}$. We will show now that we can replace
${\mc^W}_{(i)} = {\mc^W}_{(i)} \cap \overline{\S {(H_i)}}$
by the more useful formula
\begin{equation} \label{muf}
{\mc^W}_{(i)} = \overline{{\mc^W}_{(i)} \cap \S {(H_i)}}.
\end{equation}
Equation (\ref{muf}) has the advantage (over equation(\ref{3})) of requiring
only the
determination of the critical points $d \hw _{H_i} = 0$,
saving us the work of explicitly finding the ${\mc^W}_{(i)}$
points in  smaller strata. To prove (\ref{muf}) we start by taking
the closure of equation~(\ref{1}):
\begin{equation} \label{decomp}
{\mc^W}_{(i)} = \bigcup_{\S {(H)} \in \sigma_i}
\overline{\left( {\mc^W}_{(i)} \cap
\S {(H)} \right)}.
\end{equation}
Again, ${\mc^W}_{(i)}$ being irreducible means that one of the sets
in the union, say $
\overline{\left( {\mc^W}_{(i)} \cap
\S {(H_i')} \right)}$,
contains the others. To show that $\S {(H_i')} = \S {(H_i)}$
we
start from $\emptyset \neq {\mc^W}_{(i)} \cap \S {(H_i)} =
\overline{\left( {\mc^W}_{(i)} \cap
\S {(H_i')} \right)} \cap \S {(H_i)} \subseteq
 {\mc^W}_{(i)} \cap
\overline{\S {(H_i')}}  \cap \S {(H_i)}$ (here
we use that for any two sets $A$ and $B$,
$\overline{A \cap B} \subseteq \overline{A} \cap \overline{B}$.
This implies
$\overline{\S {(H_i')}}  \cap \S {(H_i)} \neq \emptyset$ and,
from equation~(\ref{int}),  $\S {(H_i')} \geq  \S {(H_i)}$.
As $\S {(H_i)}$ is the maximal set intersecting ${\mc^W}_{(i)}$
it must be $\S {(H_i')} =  \S {(H_i)}$, and equation~(\ref{muf})
follows.\\


\begin{references}

\bibitem{plb} C.~Procesi and G.~Schwarz, Phys. Lett B 161 (1985) 117.

\bibitem{gatto} R.~Gatto and G.~Sartori, Commun.~Math.~Phys~109 (1987) 327.

\bibitem{gatto2} R. Gatto and G. Sartori,
Phys. Lett. B 157 (1985) 389, Phys. Lett. B 118 (1982) 79, F. Buccella,
J.~P.~Derendinger and C.~A.~Savoy, Phys.~Lett.~B~115 (1982) 375.

\bibitem{luty} M.~A.~Luty, W.~Taylor~IV, Phys.~Rev.~D~53 (1996) 3399.

\bibitem{sc} C. Csaki, M. Schmaltz and W. Skiba, Phys. Rev. D 55 (1997) 7840,
hep-th 9612207.

\bibitem{luna} D. Luna, Bull.~Soc.~Math.~France, M\'emoire 33 (1973) 81.

\bibitem{as} M. Abud and G. Sartori, Phys. Lett. B 104 (1981) 147,
Ann. of Phys. 150 (1983) 307.

\bibitem{schwarz1} G.~Schwarz, Publ.~Math.~IHES 51 (1980) 37.

\bibitem{out} K. Intriligator, R.G. Leigh and N. Seiberg, Phys. Rev. D 50 (1994)
1092, hep-th/9403198.

\bibitem{susy} K.~Intriligator and N.~Seiberg, Nuc.~Phys~B
(Proc Suppl.) 45BC (1996) 1, hep-th/9509066;
M.E.~Peskin, TASI 96 hep-th/9702094.

\bibitem{dsb} I. Affleck, M. Dine and N. Seiberg, Nuc. Phys. B 256 (1985)
557; W. Skiba, Mod.~Phys.~Lett.A 12 (1997) 737, hep-th/9703159;
E.~Poppitz and S.~Trivedi, Phys.~Lett.B~365 (1996) 125, hep-th/9507169;
K.~Intriligator, N.~Seiberg and S.~Shenker, Phys.~Lett.B 342 (1995) 152,
hep-ph/9410203;
Y.~Shadmi and Y.~Shirman, Rev.~Mod.~Phys.~72 (2000) 25, hep-th/9907225,
and references therein.

\bibitem{clo} D.~Cox, J.~Little and D.~O'Shea, {\em Ideals, Varieties, and
Algorithms}, 2nd edition, Springer-Verlag, 1997.

\bibitem{kodaira} {\em Complex manifolds}, J. Morrow and K.~Kodaira, Holt,
Rinehart and Winston, Inc, 1971.

\bibitem{im} G.~Schwarz, Inv. Math 49 (1978), 167.

\bibitem{ela} A.G.\'Elashvili, Funct. Anal. Appl. 1 (1968), 267.

\bibitem{harris} J.~Harris, {\em Algebraic Geometry}, Springer-Verlag, 1992.

\bibitem{gv} V. Gatti (V. Kac) and E. Viniberghi (E. B. Vinberg),
Adv. Math. 30 (1978), 137.

\bibitem{sei} N. Seiberg, Nuc. Phys. B 435 (1995) 129.

\end{references}
\end{document}